\documentclass[amssymb,nobibnotes,superscriptaddress,longbibliography,notitlepage,twocolumn]{revtex4-1}
\usepackage[dvipdfmx]{graphicx}

\usepackage{enumerate,amsmath}
\usepackage{graphicx}
\usepackage{color} 

\usepackage{braket}

\begin{document}

\title{A Practical and Scalable Decoder for Topological Quantum Error Correction with Digital Annealer}%

\author{Jun Fujisaki}\email{fujisaki.jun@fujitsu.com}
\affiliation{Quantum Computing Research Center, Fujitsu Research, Fujitsu Limited., 4-1-1 Kawasaki, Kanagawa 211-8588, Japan}
\affiliation{Fujitsu Quantum Computing Joint Research Division, Center for Quantum Information and Quantum Biology, Osaka University, 1-2 Machikaneyama, Toyonaka, Osaka, 565-8531, Japan}

\author{Hirotaka Oshima}
\affiliation{Quantum Computing Research Center, Fujitsu Research, Fujitsu Limited., 4-1-1 Kawasaki, Kanagawa 211-8588, Japan}
\affiliation{Fujitsu Quantum Computing Joint Research Division, Center for Quantum Information and Quantum Biology, Osaka University, 1-2 Machikaneyama, Toyonaka, Osaka, 565-8531, Japan}

\author{Shintaro Sato}
\affiliation{Quantum Computing Research Center, Fujitsu Research, Fujitsu Limited., 4-1-1 Kawasaki, Kanagawa 211-8588, Japan}
\affiliation{Fujitsu Quantum Computing Joint Research Division, Center for Quantum Information and Quantum Biology, Osaka University, 1-2 Machikaneyama, Toyonaka, Osaka, 565-8531, Japan}

\author{Keisuke Fujii}
\affiliation{Graduate School of Engineering Science, Osaka University, 1-3 Machikaneyama, Toyonaka, Osaka 560-8531, Japan.}
\affiliation{Center for Quantum Information and Quantum Biology, Osaka University, Japan.}
\affiliation{RIKEN Center for Quantum Computing (RQC), Wako Saitama 351-0198, Japan}

\begin{abstract}
Quantum error correction is one of the most important milestones for realization of large-scale quantum computation. 
To achieve this, it is essential not only to integrate a large number of qubits with high fidelity, 
but also to build a scalable classical system that can perform error correction.
Here, we propose an efficient and scalable decoder for quantum error correction using Fujitsu Digital Annealer (DA).
Specifically, the error correction problem of stabilizer codes is mapped into an Ising-type optimization problem, so-called quadratic unconstrained binary optimization (QUBO) problem,
which is solved by DA.
In particular, we implement the proposed DA decoder for the surface code and perform detailed numerical experiments
for various code distances to see its performance and scalability. 
We observe that computational scaling for the DA decoder has a lower order of polynomial than the decoding methods using simulated annealing (SA) and minimum-weight perfect matching (MWPM) algorithm under all tested conditions.
It is also shown that the DA decoder has advantages over the Union-Find (UF) decoder from a variety of perspectives including hardware implementation.
Furthermore, the threshold behavior of the logical error probability for the DA decoder is analyzed and the resultant threshold lies between 9.4\% and 9.8\%, which is very close to that obtained by the MWPM decoder.
This result clearly shows the high potential of the DA decoder for quantum error correction.
\end{abstract}

\maketitle
\section{Introduction}
\label{sec:intro}
Quantum computers have attracted much attention because they are expected to exponentially accelerate computations in problems, such as prime factorization~\cite{shor1994}, database search~\cite{grover1996}, linear system solver~\cite{harrow2009}, and quantum chemical calculation~\cite{aspuru2005}. 
However, in order to achieve provable quantum speedup in these applications, a fault-tolerant quantum computer protected by quantum error correction is necessary~\cite{nielsen2002}. 
It has been shown that, if millions of physical qubits are realized, it will be possible to solve practically important problems that cannot be handled by classical computers~\cite{reiher2017,gidney2021}. 
While the current scale of quantum computers is still a few dozen to a hundred qubits, it is hoped that a single logical qubit can be protected from errors in the near future. 
In this direction, various experimental efforts to demonstrate quantum error correction have been reported in recent years~\cite{kelly2015state,andersen2020repeated,egan2021fault,ai2021exponential,krinner2021realizing}.

In order to realize a large-scale fault-tolerant quantum computer, the development of the quantum device itself is of prime importance, but the scalability of the classical controlling side is also an extremely challenging issue. 
In particular, decoding in quantum error correction requires error estimation from the outcomes of the syndrome measurement, which are continuously sent from the quantum device, and error correction have to be performed with low latency. 
To achieve this, quantum error correction methods using hardware of various architectures have been studied.
For example, a decoder micro-architecture that can be easily distributed 
has been developed~\cite{das2020scalable}. 
Also, a superconducting classical architecture has been proposed in which the measurement results are not transferred to room temperature, but error correction is performed in a refrigerator using a single-flux-quantum circuit~\cite{Holmes20, ueno2021qecool}.
Furthermore, to avoid the measurement and communication bottleneck, a method to perform quantum error correction with only energy dissipation and global control has been proposed by using a highly controllable classical spin system in addition to a quantum layer~\cite{fujii2014}.
Another energy-dissipative approach is the cellular automaton (CA) decoder~\cite{Herold2015, Herold2017, Kubica2018}, which can be implemented via highly parallelized integrated circuit-type hardware.
While these approaches will be further developed in future, we still need new schemes and architectures to implement them that can make good use of the scalability of today's classical computers.

In this work, we propose a decoding method for quantum error correction by using Fujitsu Digital Annealer (DA) \cite{Sao19,Matsubara20,Aramon19,DA},
which is a hardware architecture designed to solve Ising-type optimization problems, so-called quadratic unconstrained binary optimization (QUBO) problem.
The advantages of DA in terms of solving decoding problems are as follows.
First, the decoding problem of quantum error correction codes can be mapped naturally into a higher order binary optimization problem, and hence it is efficient to embed it into QUBO formulation.
This advantage also applies to other Ising solvers such as SA \cite{Kirkpatrick83} and Quantum Annealing \cite{Kadowaki98}.
Second, it solves such problems rapidly by the efficient parallel trial scheme and the massive parallelization.
Both dynamic offset and parallel tempering, also known as replica-exchange Monte Carlo \cite{Hukushima96}, increase the acceptance probability of a variable update, and also lead to speedup of the calculation.
Third, classical digital circuits constituting DA are less prone to analog noise.
In spite of these good properties and affinity between Ising model and decoding problem~\cite{fujii2014}, any Ising model solvers including DA has not been used for fast decoding in quantum error correction yet.

Since the error correction problems are defined as a binary higher order constrained optimization problem,
we map them into QUBO problems converting higher order Hamiltonian to QUBO and adding penalty terms for the constraint with providing hyperparameters. 
This allows us to solve the error correction problems by using any Ising solvers.
For error patterns generated by numerical simulations, we perform the decoding operations using DA for the planar surface code with code distances from 4 to 46.
We compare the performance of the proposed decoding method using DA with the method using SA and the MWPM algorithm \cite{Edmonds65,Galil86}, both of which are implemented on conventional CPUs.
The result shows that the computational cost of DA scales as a lower order polynomial in the number of qubits than the two approaches.
Furthermore, it is shown that the DA decoder has advantages over modern scalable approaches such as the UF decoder \cite{Delfosse17, das2020scalable} from the viewpoints of hardware implementation, applicability to the error-correcting codes, and decoding accuracy.

The rest of the paper is organized as follows.
A brief introduction to the surface code is given in Sec. \ref{sec:sc}.
The detailed formulation of correcting errors is explained in Sec. \ref{sec:procedure}.
The results of computational scaling
and the analyses of the logical error are shown in Sec. \ref{sec:analyses}.

\section{Surface code}
\label{sec:sc}
We here consider the planar surface code~\cite{Kitaev2003,Bravyi1998,Fowler12}, which is considered to have high experimental feasibility.
However, the following argument can be generalized straightforwardly for an arbitrary stabilizer code.
Below we will briefly explain quantum error correction using the surface code with fixing our notations.
Let us consider a square lattice shown in Fig.~\ref{fig2}, where a data qubit is located on each edge shown by a red square. 
The linear length of the square lattice is called the code distance.
The stabilizer operators are defined as the products of Pauli-$Z$ and $X$
operators associated with each face $f$ and vertex $v$, 
respectively, as follows:
\begin{align}
    A_f  &= \prod _{i \in \partial f } Z_i ,
    \label{Z-stabilizer}
    \\
    B_v &= \prod _{j \in \delta v} X_j,
    \label{X-stabilizer}
\end{align}
where $\partial f$ and $\delta v$ indicate
sets of edges surrounding face $f$ and incident to vertex $v$, respectively.
The surface code state $|\Psi \rangle$
is defined as a simultaneous eigenstate of the stabilizer operators with eigenvalue $+1$:
\begin{align}
    A_f |\Psi \rangle &  = |\Psi \rangle 
    \textrm{ for all } A_f,
    \label{Z-eigen-eq}
    \\
    B_v |\Psi \rangle & = |\Psi \rangle 
    \textrm{ for all }  B_v.
\end{align}

Suppose a Pauli operator $P$ occurs on the code state, the eigenvalues with respect to the stabilizer operators that anti-commute with the error $P$ are flipped from $+1$ to $-1$. 
Since this provides information about errors, the set of measured eigenvalues is called error syndrome.
For simplicity, we assume $X$ and $Z$ errors occur independently with probability $p$ for each qubit. 
This allows us to consider $X$- and $Z$-errors separately, and we can discuss only $Z$-errors, which can be detected by $X$-type syndrome measurements done on each vertex.

Let us denote the Pauli-$Z$ operators acting on qubits defined on a subset $E$ of edges as $Z(E)$.
Such an error is detected by odd eigenvalues $-1$ at the boundary $\partial E$ of the error chain $E$, i.e., 
\begin{align}
    B_v Z(E) |\Psi \rangle = - Z(E) |\Psi \rangle ,
\end{align}
iff $v \in \partial E$.
This means that the obtained $X$-type syndrome $S$ corresponds to the boundary $\partial E$
of the (unknown) error chain $E$,
i.e., $S=\partial E$.
Therefore, 
the error correction problem in the surface codes is to find an error chain $E$ that satisfies the boundary condition from the information in the boundary.
Since we want to find the most likely error for the given syndrome $S$, the decoding process can be
written as
\begin{align}
    E^* = \arg \max _{E}  {\rm Prob}(E|S=\partial E) .
    \label{solution}
\end{align}
Since we assume that errors are located independently with probability $p$, 
this reads
\begin{align}
    {\rm min} |E| 
    \textrm{  s.t. } E = \partial S,
\end{align}
where $|E|$ indicates the number of Pauli-$Z$ operators in $E$.
This is an optimization problem formulated as an integer programming
and the above argument can be applicable 
for an arbitrary stabilizer code.
Specifically, for the surface code, 
this task is interpreted as finding the shortest chains that connects two odd eigenvalues, and hence can be solved efficiently 
by using the MWPM algorithm.
However, the MWPM algorithm is too complicated to be implemented on domain-specific architecture such as FPGAs. Furthermore,
it is not applicable to more general Pauli errors nor general stabilizer codes.
This is the main reason why we want to apply DA, a hardware architecture for the decoding problem.

\section{Ising model formulation of error correction}
\label{sec:procedure}

We describe a specific procedure for mapping an error correction problem in the surface code into an Ising-type optimization problem.
Here the errors are represented by Ising spins, where if an error occurs on a qubit, the corresponding spin $\sigma$ is flipped from $+1$ to $-1$.
Note that we treat only Pauli-$Z$ errors, because the Pauli-$X$ errors can be handled in the same way by introducing spins for them.
Let $b_v$ be a syndrome corresponding to the $X$-type stabilizer $B_v$.
Then the Ising Hamiltonian for the error correction is defined by the interaction term with 4 spins and the external field term:
\begin{eqnarray}
H=-J\sum^{N_{v}}_{v}b_{v}\prod^{4}_{i\in \delta{v}}\sigma_i-h\sum^{N_{d}}_{i}\sigma_{i}
\label{eq1},
\end{eqnarray}
where $J$ and $h$ are the parameters
as explained later, $N_v$ and $N_{d}$ are the numbers of X-type stabilizer operators and data qubits, respectively, $\sigma_{i}$ is the $i$-th spin variable. 
The first term of Eq.~(\ref{eq1}) with four-body interactions imposes that errors have to satisfy the given error syndrome,
where $J$ is a hyperparameter for this constraint.
The second term of the external field minimizes the number of errors, which can be controlled by the hyperparameter $h$.
Thereby,
the global energy minimum state configuration corresponds to the most likely error satisfying the syndrome condition in the limit of large $J$.

\subsection{Designing the cost function}
\label{sec:cost function}

From error syndromes obtained by measurements, Ising Hamiltonian for decoding with DA, namely cost function, is prepared.
Because Eq.~(\ref{eq1}) has the four-body interaction and thus is a higher order binary optimization problem, it is not possible for DA to process the function with its original form.
Converting it into a QUBO problem, is done as follows \cite{Ishikawa11,Xia18}.

First, Ising spin $\sigma_{i}$ is converted into the binary variable $x_{i}$ that can be handled in a digital circuit by
\begin{equation}
x_{i}=\frac{\left(1-\sigma_{i}\right)}{2}
\label{eq4}.
\end{equation}
Then, since Eq.~(\ref{eq1}) contains up to fourth order terms for $x_{i}$, the following auxiliary binary variables $\{ z_k \}$ are introduced to represent them in up to second order terms:
\begin{equation}
z_{k}=x_{i}x_{j} \;\;\; (0 \leq k < N_{m})
\label{eq5},
\end{equation}
where $N_{m}$ is the number of the auxiliary binary variables.
In order to impose the above equality, the following penalty term is further required:
\begin{equation}
H_{\rm penalty}=\alpha\sum^{N_{m}}_{m}\left[x_{i}x_{j}-2z_{m}\left(x_{i}+x_{j}\right)+3z_{m}\right]
\label{eq6},
\end{equation}
where $\alpha$ is the hyperparameter to control the penalty term and is set to $8J$ in our study.
Note that $H_{\rm penalty}$ takes a positive value if Eq.~(\ref{eq5}) is not satisfied.

In this way, we obtain 
the QUBO Hamiltonian $H'$ consisting of the original Hamiltonian $H$ and additional penalty term $H_{\rm penalty}$,
\begin{equation}
H'=H+H_{\rm penalty}\\=-\frac{1}{2}\sum^{N}_{i,j}W_{ij}y_{i}y_{j}-\sum^{N}_{i}V_{i}y_{i}+c
\label{eq2},
\end{equation}
where $W_{ij}$ is the weight coefficient calculated from $J$ and $\alpha$, and $V_{i}$ is the bias coefficient calculated from $J$, $h$, and $\alpha$.
$c$ is the constant number.
For more details on these relationships between parameters, see Appendix \ref{sec:detailed cost function}.

Note that $y_{i}$ is defined such that
\begin{equation}
y_{i} = 
\begin{cases}
x_{i} & (0 \leq i < N_{d})\\
z_{i} & (N_{d} \leq i < N_{d}+N_{m})
\end{cases}
\label{eq3},
\end{equation}
for simplicity of notation. 
$N$ indicates the total number of the binary variables,
\begin{equation}
N=N_{d}+N_{m}.
\end{equation}
The initial values of all the binary variables are set to zero in every calculation.

\begin{figure*}
\centering
\includegraphics[scale=0.7]{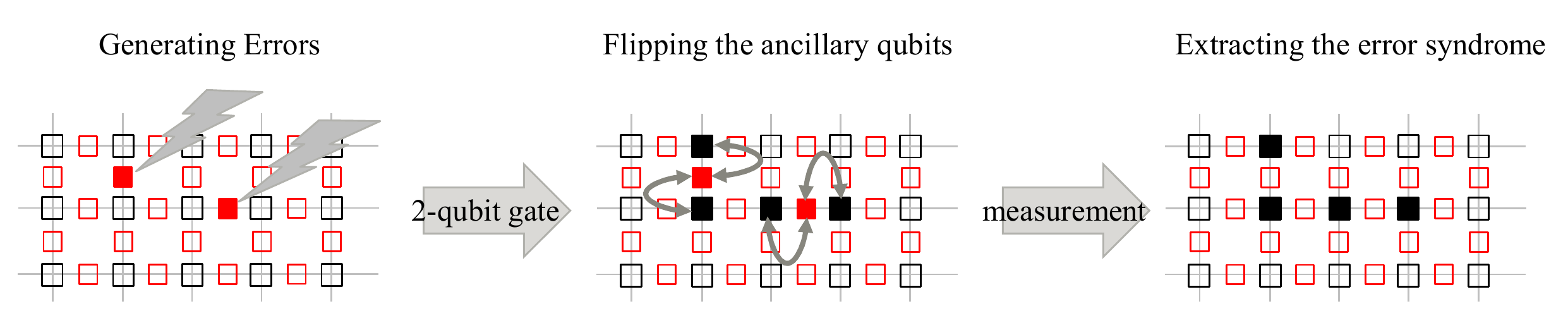}
\caption{\label{fig:wide}Conceptual diagram of error syndrome extraction.
Here, we focus on the Z error, and only X-type ancillary qubits are depicted as the open squares (black) on the nodes.
The data qubits and the flipped data qubits are depicted as the open squares (red) and the filled squares (red) on the edges, respectively. These processes are operated by simulation on a classical computer instead of a quantum computer.}
\label{fig2}
\end{figure*}

\subsection{DA for decoding}
\label{subsec:DA}

There are several ways to solve QUBO such as SA and Quantum Annealing.
Specifically, we employ DA \cite{Sao19,Matsubara20,Aramon19,DA}, which is a hardware architecture as a solver for the QUBO problems that have the cost function described above.
DA can solve such problems rapidly by the efficient parallel trial scheme and the massive parallelization.
Furthermore, DA can handle complicated problems accurately due to its all-to-all connection architecture and noise tolerance which is characteristic of classical digital circuits.
All calculations are performed on the second-generation DA environment prepared for research use \cite{Matsubara20}.

DA iterates, for a fixed number of times, searching for a binary variable $x_{i}$ whose update decreases the total energy in Eq.~(\ref{eq2}) or satisfies the acceptance condition in Metropolis criterion \cite{Kirkpatrick83}.
Although the latter case of update increases the total energy, it plays an important role for the system to escape from the local energy minimum.
Each time such a binary variable is found, DA updates it and continues with the subsequent steps.
In the calculation, we record the number of iterations and the state of all the binary variables when the minimum value of the total energy is updated.

The positions where binary variables $x_{i}$ are equal to 1 obtained from the calculation are regarded as the positions of the actual errors.
Ideally, the obtained state is the global energy minimum state.
However, the state might not be in the global minimum depending on the error pattern due to the failure of escaping from the local energy minimum state.
Even in such cases, under certain conditions we are able to specify recovery operators for error correction.
The reason for this is described below.

In the surface code, if the estimated and the actual error positions form trivial loops, they can be represented by products of stabilizer operators defined in Eq.~(\ref{Z-stabilizer}).
Because a logical qubit state is a simultaneous eigenstate of these stabilizer operators as shown in Eq.~(\ref{Z-eigen-eq}), such trivial loops do not harm the logical qubit state.
If they form an end-to-end chain corresponding to a logical operator, on the other hand, it causes the logical error, and the error correction fails.
In Appendix \ref{sec:demo}, we show that our DA decoder can correct errors in principle by using two examples with different error patterns.

\section{Detailed analyses of DA decoder}
\label{sec:analyses}

\subsection{Computational scaling and comparison with other methods}
\label{sec:scaling}

From a practical point of view, 
decoding should be efficient, and obtained solutions of Eq.~(\ref{solution}) must keep syndrome constraints for many qubits with various error patterns.
For this reason, in this section, we perform exhaustive calculations 
and evaluate the accuracy and the computational scaling of error correction over a wide range of code distances under several physical error rates $p$.
In addition, the computational scaling is compared with those obtained with SA, which is often used to solve Ising-type optimization problems, and an
MWPM algorithm, which is widely used as a decoder for the surface code.

\begin{table}[b]
\caption{\label{tab:table2}%
The set of parameters for the DA decoder used in the exhaustive survey.
}
\begin{ruledtabular}
\begin{tabular}{ll}
\textrm{Parameter}&
\textrm{Value}\\
\colrule
Number of data qubits $N_d$ ( code distance $d$ ) & 25--4141 ( 4--46 )\\
Physical error rate $p$ & 0.1--20\%\\
$J$ & 1024\\
$h$ & 1\\
Annealing mode & Replica exchange\\
Number of replicas & 128\\
Maximum temperature & 5\\
\end{tabular}
\end{ruledtabular}
\end{table}

In the survey, we simulate syndrome measurement process classically because real quantum devices with hundreds or thousands of qubits do not exist yet.
As shown in Fig.~\ref{fig2}, we randomly generate errors on data qubits and calculate their parity according to the stabilizers.
The total number of data qubits is varied from 25 ( code distance $d$ = 4 ) to 4141 ( $d$ = 46 ), with 
$p$ being 0.1\%, 1.0\%, 2.0\%, 5.0\%, 10\%, and 20\%. The parameter sets are listed in TABLE \ref{tab:table2}. 
The probability of syndrome constraint solution
and the number of iterations for the final update of minimization in total energy are evaluated for 100 error patterns for each parameter set. 

\begin{figure}[h]
\includegraphics[width=8.6cm, keepaspectratio]{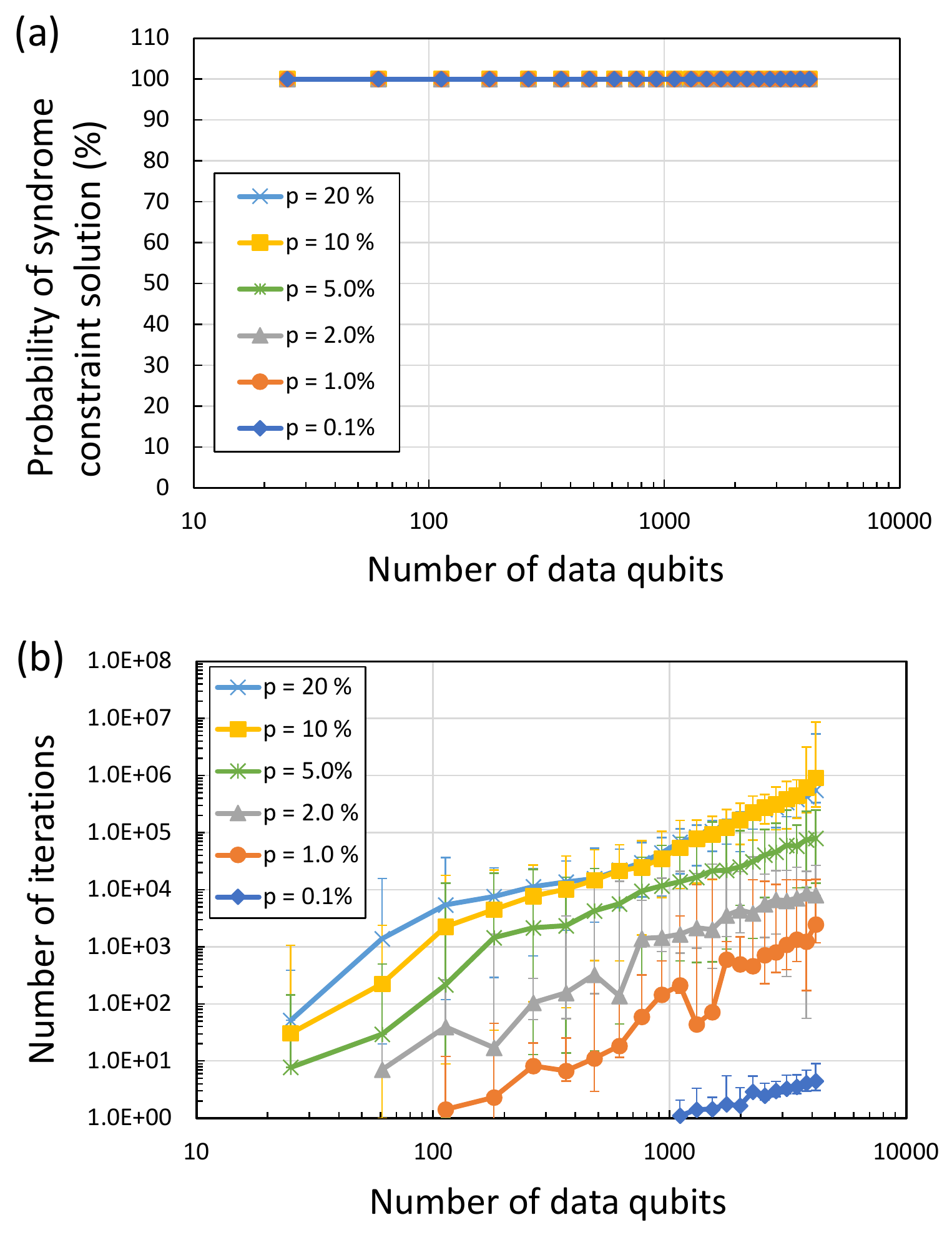}
\caption{\label{fig:DAresults}Results of decoding with DA.
(a) The probability of syndrome constraint solution is always 100\% for any condition.
(b) The average number of iterations for each number of data qubits and physical error rate.}
\label{fig4}
\end{figure}

The results are shown in Fig.~\ref{fig4}.
The probability of syndrome constraint solution
is 100\% for all the cases, as shown in Fig.~\ref{fig4}(a).
This indicates that robust decoding is possible using DA.
The average values of the numbers of iterations 
are shown in the double logarithmic plots in Fig.~\ref{fig4}(b).
The error bars show the minimum and the maximum numbers of iterations for each data point.
Since each plot is approximately linear in a double logarithmic plot, the number of iterations appears to scale as a polynomial function of the number of qubits.
Regression analysis over the entire domain shows that the degree of the polynomial is 1.01 (minimum) for $p = 0.1\%$ and 1.84 (maximum) for $p = 5.0\%$. All data are presented in TABLE \ref{tab:table3}.

\begin{figure}[h]
\includegraphics[width=8.6cm, keepaspectratio]{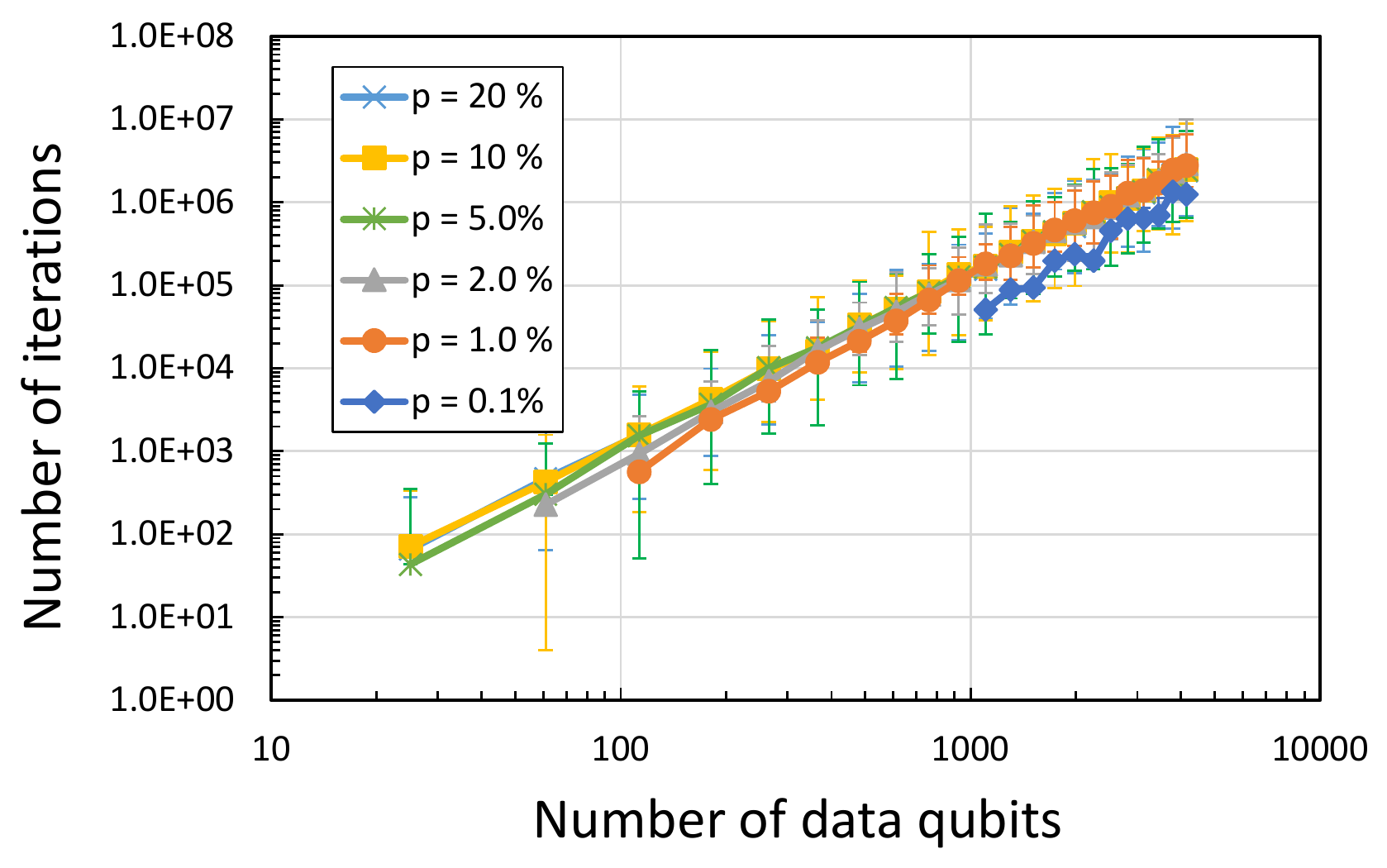}
\caption{\label{fig:SAresult}Results of the SA decoder with the same parameter values for DA.
In contrast to Fig.~\ref{fig4}(b), where $p$ is small, the number of iterations increases significantly.}
\label{fig5}
\end{figure}

To compare with other decoding methods, SA and MWPM are then considered.
For the SA decoder, the parameter values of the Ising Hamiltonian Eq.~(\ref{eq1}) and maximum temperature are the same as those for the DA decoder.
It is confirmed that 
the probability of syndrome constraint solution of each parameter is always 100\%, and the numbers of iterations in decoding are shown in Fig.~\ref{fig5}. 

\begin{table}[b]
\caption{\label{tab:table3}Order of polynomial $n$ for the fixed physical error rates of the DA, SA, and MWPM decoders.}
\begin{ruledtabular}
\begin{tabular}{cccc}
 Physical error rate $p$ (\%)& $n$ for DA & $n$ for SA & $n$ for MWPM \\ \hline
 0.1 & 1.01 & 2.77 & 2.06\\
 1.0 & 1.79 & 2.38 & 2.10\\
 2.0 & 1.74 & 2.25 & 2.24\\
 5.0 & 1.84 & 2.11 & 2.50\\
 10 & 1.81 & 2.05 & 2.72\\
 20 & 1.54 & 2.04 & 2.83\\
\end{tabular}
\end{ruledtabular}
\end{table}

With SA, the number of iterations also scales polynomially.
As shown in TABLE \ref{tab:table3}, the degree of the polynomial for each 
$p$ is between 2.04 and 2.77, which is greater than that with DA.
In particular, the numbers of iterations for lower 
$p$ (0.1\%--2.0\%) with SA (Fig.~\ref{fig5}) are much greater than those with DA (Fig.~\ref{fig4}(b)).
This is arguably due to the difference in search algorithms.
In SA, because candidate binary variables for update are randomly selected and only a single variable is treated in each iteration, a large number of iterations are required to find a solution even for a lower 
$p$. In contrast, thanks to the parallel-trial scheme and the massive parallelization, the number of iterations in DA becomes quite low for lower 
$p$.
\begin{figure}[h]
\includegraphics[width=8.8cm, keepaspectratio]{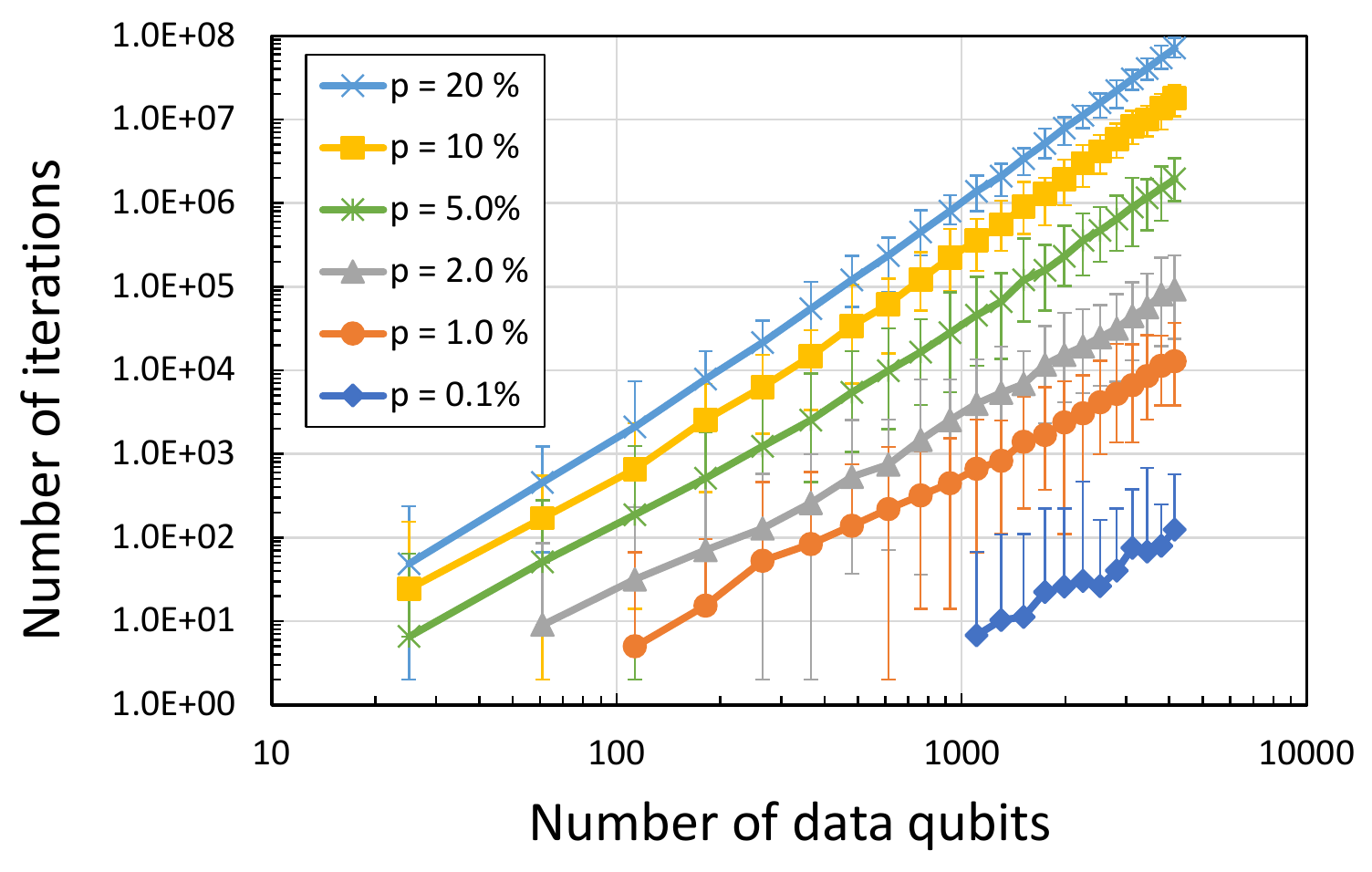}
\caption{\label{fig:MWPMresult}Results of the MWPM decoder.
For large physical error rates, the number of iterations increases almost in proportion to the cube of the number of data qubits, which agrees with the theoretical value.}
\label{fig6}
\end{figure}

For an MWPM decoder, we use an open source software (NetworkX \cite{Networkx}) for Edmonds' blossom algorithm \cite{Edmonds65}.
The probability of syndrome constraint solution
for the MWPM decoder is confirmed to be 100\% by its definition, similarly to that of the DA and SA decoders. 
The number of iterations in MWPM is counted as the number of edge searches in the matching \cite{Galil86}, and the results are shown in Fig.~\ref{fig6}. 
Because the number of iterations in the MWPM decoder is not directly comparable to the DA and SA decoder, we focus only on scaling.
The amount of computation is apparently scaled in a polynomial manner, and the degree of the polynomial is 2.83 at 
$p = 20\%$. This value is nearly equivalent to the theoretical value of the Edmonds' blossom algorithm, which has $\mathcal{O}\left(N^{3}\right)$ scaling.
When $p$ is 0.1\%, the exponent is a little less than the theoretical value, but it is still above 2.
For a detailed discussion of the scalability difference between the DA and MWPM decoder, see Appendix \ref{sec:characteristics}.

We should note that we are focusing in this study on the accuracy and scalability of the decoding algorithms. While actual time required for decoding is also important for practical use, we do not compare or optimize the actual time here, since the actual time required for SA or MWPM varies greatly depending on the architecture (or CPUs) in which they are implemented.
Although the current version of DA is not designed to provide the exact processor time required for a single optimization calculation, it is roughly estimated to be on the order of one microsecond for one thousand qubits. This is promising compared to other decoding algorithms and worth further study.

\subsection{Logical error rate}
\label{sec:logical}

One of the most important indices to evaluate the capability of decoders for practical use is the logical error rate.
Here we evaluate the error threshold by performing decoding with DA for the code distance $d$ of 5, 11, 21, 31, and 41.
The threshold is the value of the physical error rate below which the logical error rate can be suppressed arbitrarily by extending the code distance. 
For each physical error rate, 10,000 samples were calculated to determine the logical error rate.

\begin{figure}[h]
\includegraphics[width=9cm, keepaspectratio]{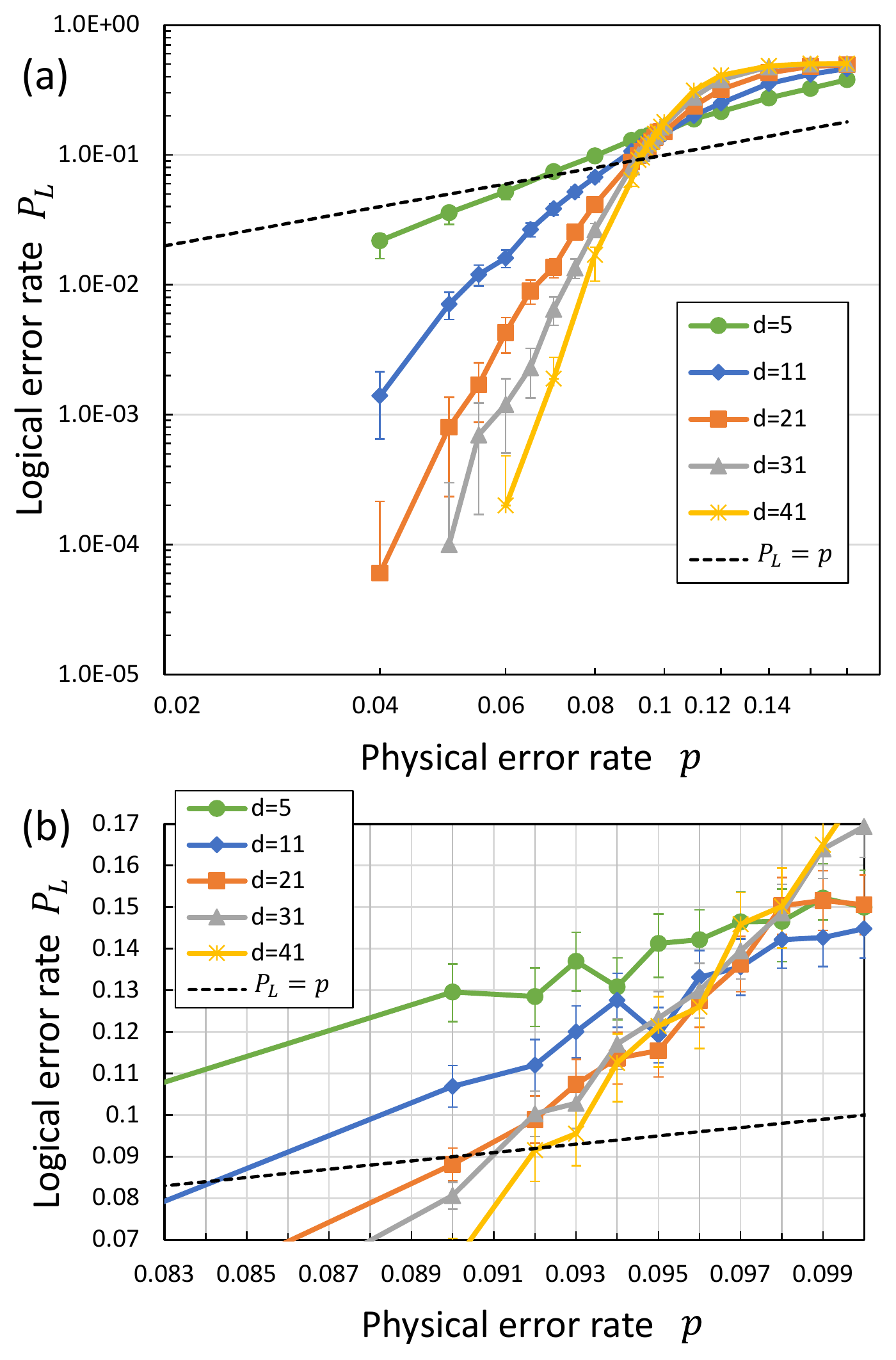}
\caption{\label{fig:logical}Correlation between the calculated logical error rate $P_{L}$ and the physical error rate $p$ for 
$d$ = 5, 11, 21, 31, and 41.
The error bars indicate the standard errors.
(a) $P_{L}$ for a wide range of $p$. (b) Enlarged view of the vicinity of the threshold.}
\label{fig10}
\end{figure}

The results are shown in Fig.~\ref{fig10}.
From Fig.~\ref{fig10}(a), we confirmed that the DA decoder shows appropriate behavior where a threshold is estimated by evaluating logical errors for various $d$.
As shown in Fig.~\ref{fig10}(b), the threshold for the DA decoder lies between 9.4\% and 9.8\%.
The slight difference from those with the MWPM decoder \cite{Fowler12} is probably due to the difference in search algorithms or the temperature schedule in DA.
However, the high values of the thresholds obtained here show evidence of the capability of the DA decoder. 

In addition to the thresholds, the parameters $c_{1}$ and $c_{2}$ in the following power-law equation \cite{Fowler12} are derived by fitting the logical error rates calculated with $p$ between 4\% and 8\%.
\begin{equation}
P_{L}=c_{1}\left( \frac{p}{p_{\rm th}} \right)^{c_{2}d_{e}}
\label{eq8},
\end{equation}
where, $P_{L}$ is the logical error rate, $c_{1}$ and $c_{2}$ are the parameters, $p_{\rm th}$ is the threshold, and $d_{e}$ is defined by the following floor function,
\begin{equation}
d_{e}=\left \lfloor \frac{d+1}{2} \right \rfloor
\label{eq9}.
\end{equation}
The values for $c_{1}$ and $c_{2}$ for each $d$ are listed in TABLE \ref{tab:table5}.
For small $d$, $c_{2}$ is close to 1.0 and matches with the value of the MWPM decoder.
For example, at $p=0.1\%$, if this scaling is correct, then at $d=11$ the logical error rate can be greatly reduced to $4\times10^{-11}$.
These results indicate that the DA decoder has a promising capability as a decoder for quantum error correction.\\

\begin{table}[t]
\caption{\label{tab:table5}%
Parameter fit results. $c_{1}$ and $c_{2}$ are the fitting parameters of the suppression law of logical error rates.
}
\begin{ruledtabular}
\begin{tabular}{lccr}
\textrm{Fitting parameter}&
\textrm{$d=11$}&
\textrm{$d=21$}&
\textrm{$d=31$}\\
\colrule
$c_{1}$ & 0.18 & 0.23 & 0.24\\
$c_{2}$ & 0.81 & 0.77 & 0.70\\
\end{tabular}
\end{ruledtabular}
\end{table}

\subsection{Comparison with other decoders}
\label{sec:comparison}

The results presented in 
\ref{sec:scaling} indicates that the DA decoder is more scalable than the SA and MWPM decoders.
In addition to these methods, there are other approaches to the scalable decoding such as the renormalization group (RG) decoder \cite{Duclos-Cianci2010a, Duclos-Cianci2010b}, the CA decoder, and the UF decoder. 
The RG decoder has computational scaling $\mathcal{O}\left(N \mathrm{log} N\right)$ \cite{Duclos-Cianci2010a}, which is slightly larger than the DA decoder.
Moreover, its threshold value in code capacity noise is comparable to the DA decoder.
The computational scaling of the CA decoder is polylogarithmic \cite{Herold2015}.
However, the threshold value is even smaller (8.2\%), and it requires highly optimized hardware and is not yet implemented.
In this respect, the DA decoder is already implemented in actual hardware architecture and has a higher threshold than those decoders.

The UF decoder has almost linear computational scaling, $\mathcal{O}\left(N \alpha(N) \right)$ \cite{Delfosse17}. 
The computational scaling of the DA decoder is comparable to the UF decoder at $p$ = 0.1\%, but the former degrades at large $p$.
However, the DA decoder has three advantages over the UF decoder.
First, the UF decoder has relatively complex algorithm from viewpoint of hardware implementation, such as cluster expansion and collision separation/fusion processing.
On the other hand, the DA decoder can be implemented with very simple algorithm such as optimization of the Ising Hamiltonian.
Second, the DA decoder precedes the UF decoder not only in theoretical proposals and virtual testing on general-purpose CPUs, but also in hardware implementation and testing using actual dedicated architecture.
For the UF decoder, at present, there are no hardware implementations other than general-purpose CPUs, such as FPGAs and ASICs.
Third, while the DA decoder can be applied to arbitrary stabilizer codes, the UF decoder can be applied to a certain restricted class of quantum codes.
Moreover, when comparing the logical error rate without measurement errors, the values of the UF decoder are 
14\% or more when code distance d is 34 as shown in Fig. 6 in \cite{Delfosse17}. Those values are higher than that of the DA decoder with d = 11 in Fig.~\ref{fig10}(b).

\section{Conclusion}
\label{sec:conclusion}

In this paper, we showed that practical and scalable error correction in the surface code can be achieved with the DA decoder.
The significant advantage of the DA decoder is that the computational scaling is $\mathcal{O}\left(N^{1.01\thicksim1.84}\right)$, and the polynomial order of scaling with the DA decoder is smaller than that with the SA and MWPM decoders under all the tested conditions.
It should be noted that the order is almost linear when $p$ = 0.1\%.
We also note that the DA decoder is expected to be applicable in principle to other topological codes, such as color codes where the MWPM algorithm cannot be directly applied \cite{Bombin13}.
These facts suggest that the error correction architecture with high-performance Ising hardware, such as DA, is a promising approach for scalable error correction systems in the future. 

We mainly discussed decoding on the assumption that measurements are perfect because this is the first step toward the establishment of the DA decoder.
However, imperfect syndrome measurement can occur in practice, so analysis including measurement errors is of great importance.
If we consider measurement errors, multiple syndrome measurements will be necessary and the expression of the interaction in the Hamiltonian needs to be modified.
We believe that this extension is possible and we will solve the same type of combinatorial optimization problem discussed in this paper.
Further improvement and extension to more realistic noise models is an intriguing future work for the DA decoder.

In our analysis, we imposed a 100\% 
syndrome constraint solution
for error correction, 
that is, the estimated errors have to satisfy the syndrome condition.
However, this would make the corresponding optimization problem harder, 
making the number of iterations larger.
We could relax this condition so that the residual errors which cannot be corrected 
in the previous round are corrected in the following round
as done in Ref.~\cite{fujii2014}. 
In such a case, the scalability may be improved by reducing the number of iterations.
Besides, if we think quantum error correction with imperfect syndrome measurements,
the errors are not necessarily corrected within a single round of the syndrome measurement.

There are still various implementation issues that commonly exist for the realization of such quantum-classical hybrid systems ~\cite{das2020scalable,Holmes20,ueno2021qecool}.
In particular, the latency constraints for the data communication between quantum and classical computers and the extra processing in classical computers are key issues.
For these purposes, it is necessary to carry out comprehensive research and development from the viewpoint of the whole computer architecture. 

\section*{Acknowledgement}
We would like to thank Kazuya Takemoto, Toshiyuki Miyazawa, Yoshinori Tomita, Noboru Yoneoka, Kentaro Katayama, and Tomochika Kurita for their technical support in using Digital Annealer. We would also like to thank Daisuke Kushibe, Tatsuya Sakashita, Yusaku Takeuchi, Mitsuki Katsuda, Hideaki Hakoshima, Hiroshi Ueda, and Kosuke Mitarai for their helpful discussions. KF is supported by MEXT Quantum Leap Flagship Program (MEXT Q-LEAP) Grant No. JPMXS0118067394 and JPMXS0120319794, JST COI-NEXT Grant No. JPMJPF2014, and JST Moonshot R\&D Grant No. JPMJMS2061.

\appendix

\section{Concrete formula for the cost function}
\label{sec:detailed cost function}

We here describe the details of the cost function explained in \ref{sec:cost function}. 
The original form of the Hamiltonian to be minimized is defined by the interaction term with 4 spins $\sigma_i$ and the external field term as shown Eq.~(\ref{eq1}).

First, after the spin variable $\sigma_i$ is converted into a binary variable $x_{i}$ according to the equation Eq.~(\ref{eq4}), Eq.~(\ref{eq1}) is transformed as follows

\begin{equation}
\begin{split}
H &=-16J\sum^{N_{v}}_{v}b_{v}x_{i}x_{j}x_{k}x_{l}\\
  &+8J\sum^{N_{v}}_{v}b_{v}\left(x_{i}x_{j}x_{k}+x_{i}x_{j}x_{l}+x_{i}x_{k}x_{l}+x_{j}x_{k}x_{l}\right)\\
  &-4J\sum^{N_{v}}_{v}b_{v}\left(x_{i}x_{j}+x_{i}x_{k}+x_{i}x_{l}+x_{j}x_{k}+x_{j}x_{l}+x_{k}x_{l}\right)\\
  &+2J\sum^{N_{v}}_{v}b_{v}\left(x_{i}+x_{j}+x_{k}+x_{l}\right)\\
  &-J\sum^{N_{v}}_{v}b_{v}+2h\sum^{N_{d}}_{i}x_{i}-h\sum^{N_{d}}_{i}.
\label{eqH2}
\end{split}
\end{equation}

Next, using the auxiliary variable $z_k$ defined in the equation Eq.~(\ref{eq5}), the above expression becomes the Hamiltonian in QUBO format

\begin{equation}
\begin{split}
H &=-16J\sum^{N_{v}}_{v}b_{v}z_{m}z_{n}\\
  &+8J\sum^{N_{v}}_{v}b_{v}\left(z_{m}x_{k}+z_{m}x_{l}+x_{i}z_{n}+x_{j}z_{n}\right)\\
  &-4J\sum^{N_{v}}_{v}b_{v}\left(x_{i}x_{j}+x_{i}x_{k}+x_{i}x_{l}+x_{j}x_{k}+x_{j}x_{l}+x_{k}x_{l}\right)\\
  &+2J\sum^{N_{v}}_{v}b_{v}\left(x_{i}+x_{j}+x_{k}+x_{l}\right)\\
  &-J\sum^{N_{v}}_{v}b_{v}+2h\sum^{N_{d}}_{i}x_{i}-h\sum^{N_{d}}_{i}.
\label{eqH3}
\end{split}
\end{equation}

Finally, the cost function optimized by the DA is obtained by adding the penalty term defined in the equation Eq.~(\ref{eq6}).

\begin{equation}
\begin{split}
H' &=-16J\sum^{N_{v}}_{v}b_{v}z_{m}z_{n}\\
  &+8J\sum^{N_{v}}_{v}b_{v}\left(z_{m}x_{k}+z_{m}x_{l}+x_{i}z_{n}+x_{j}z_{n}\right)\\
  &-4J\sum^{N_{v}}_{v}b_{v}\left(x_{i}x_{j}+x_{i}x_{k}+x_{i}x_{l}+x_{j}x_{k}+x_{j}x_{l}+x_{k}x_{l}\right)\\
  &+2J\sum^{N_{v}}_{v}b_{v}\left(x_{i}+x_{j}+x_{k}+x_{l}\right)\\
  &-J\sum^{N_{v}}_{v}b_{v}+2h\sum^{N_{d}}_{i}x_{i}-h\sum^{N_{d}}_{i}\\
  &+\alpha\sum^{N_{m}}_{m}\left[x_{i}x_{j}-2z_{m}\left(x_{i}+x_{j}\right)+3z_{m}\right].
\label{eqH4}
\end{split}
\end{equation}

By comparing this equation with the equation Eq.~(\ref{eq2}), the coefficients $W_{ij}$ and $V_{i}$ for a binary variable $y_{i}$ can be related to parameters such as $b_{v}$, $J$, $h$, and $\alpha$.

\section{Demonstrations of correcting errors by DA}
\label{sec:demo}
 
To show that the DA decoder works with sufficient performance, we demonstrate here two examples of parameter settings
as listed in in TABLE \ref{tab:table1}.
Note that for sample 1, the code distance is chosen to be 6, and the total number of data qubits is 51. For sample 2, they are 41 and 3281, respectively.
For Sample 1, the physical error rate is deliberately set as high as 20\% to increase the number of errors in order 
to illustrate how errors are corrected, while it is apparently above the threshold.

\begin{table}[b]
\caption{\label{tab:table1}%
Set of parameters for the DA decoder used in the demonstration.
}
\begin{ruledtabular}
\begin{tabular}{lll}
\textrm{Parameter}&
\textrm{Value for sample 1}&
\textrm{Value for sample 2}\\
\colrule
Number of data qubits $N_d$ & 51 & 3281\\
Physical error rate $p$ & 20\% & 2.0\%\\
$J$ & 4 & 4\\
$h$ & 1 & 1\\
Annealing mode & Replica exchange &Replica exchange\\
Number of replicas & 128 & 128\\
Maximum temperature & 10 & 10\\
\end{tabular}
\end{ruledtabular}
\end{table}
The values of $J$ and $h$ are determined by whether or not there is an energy gain in Eq.~(\ref{eq1}) when multiple spins are flipped.
More specifically, if the length of the error chain is $n$, the corresponding energy change due to spin flipping is $-4J+2nh$.
It must be less than zero for such flips to be realized, which leads to the following constraint,
\begin{equation}
J>\frac{n}{2}h
\label{eq7}.
\end{equation}
For example, if $n$ is 6 and $h$ is 1, $J$ must be greater than 3 for the error chain to be corrected.
Therefore, the values of the parameters $J$ and $h$ in both the samples are set to 4 and 1, respectively, assuming that the number of connected errors is less than or equal to 6.

\begin{figure}[h]
\includegraphics[scale=0.7]{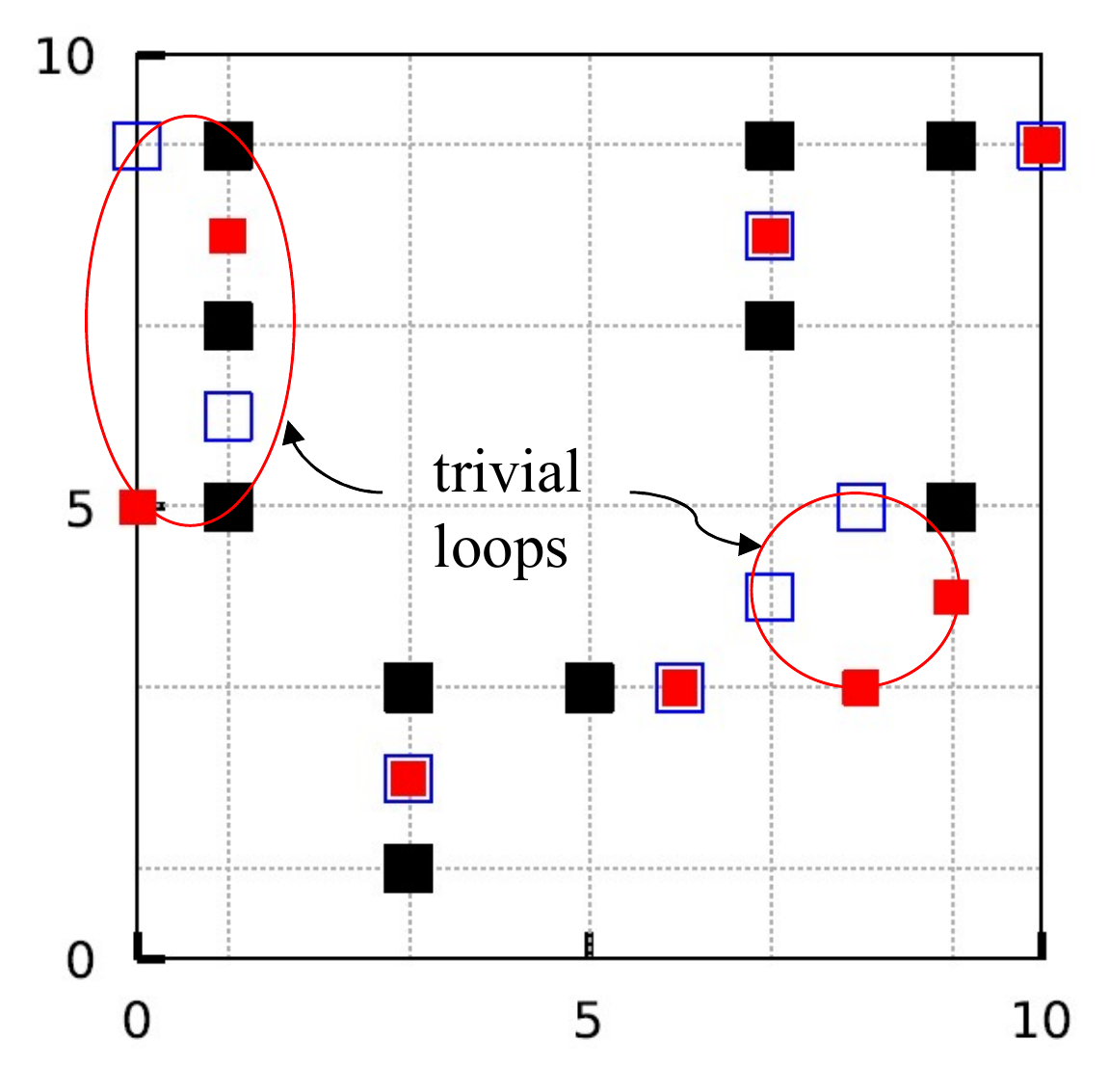}
\caption{\label{fig:sample1}Error correction result with the DA decoder for sample 1.
The filled small squares (red) indicate the actual Z errors, and the open squares (blue) indicate the corrected errors.
Some of them are in the identical positions, and the rest constitute trivial loops that do not harm the logical qubit state.}
\label{fig3}
\end{figure}

As an illustrative example, the result obtained with the above parameter set of sample 1 is shown in Fig.~\ref{fig3}. 
The filled small squares (red), the filled large squares (black), and the open squares (blue) indicate the data qubits with errors, vertices ($X$-type stabilizers) with $b_v=-1$, and the estimated error locations obtained as a solution of DA, respectively.
The other qubits are not drawn for ease of viewing.
Decoding by DA, as mentioned earlier, uses only information about the error syndrome (filled large squares).
The syndrome constraint solution is obtained in Fig.~\ref{fig3}
because half of the actual errors and the corrected errors coincide, and the others form trivial loops.
In this case the total energy of this system matches that with the actual errors.
Thus, one of the most likely errors is obtained by the DA decoder.

\begin{figure}[h]
\includegraphics[scale=0.8]{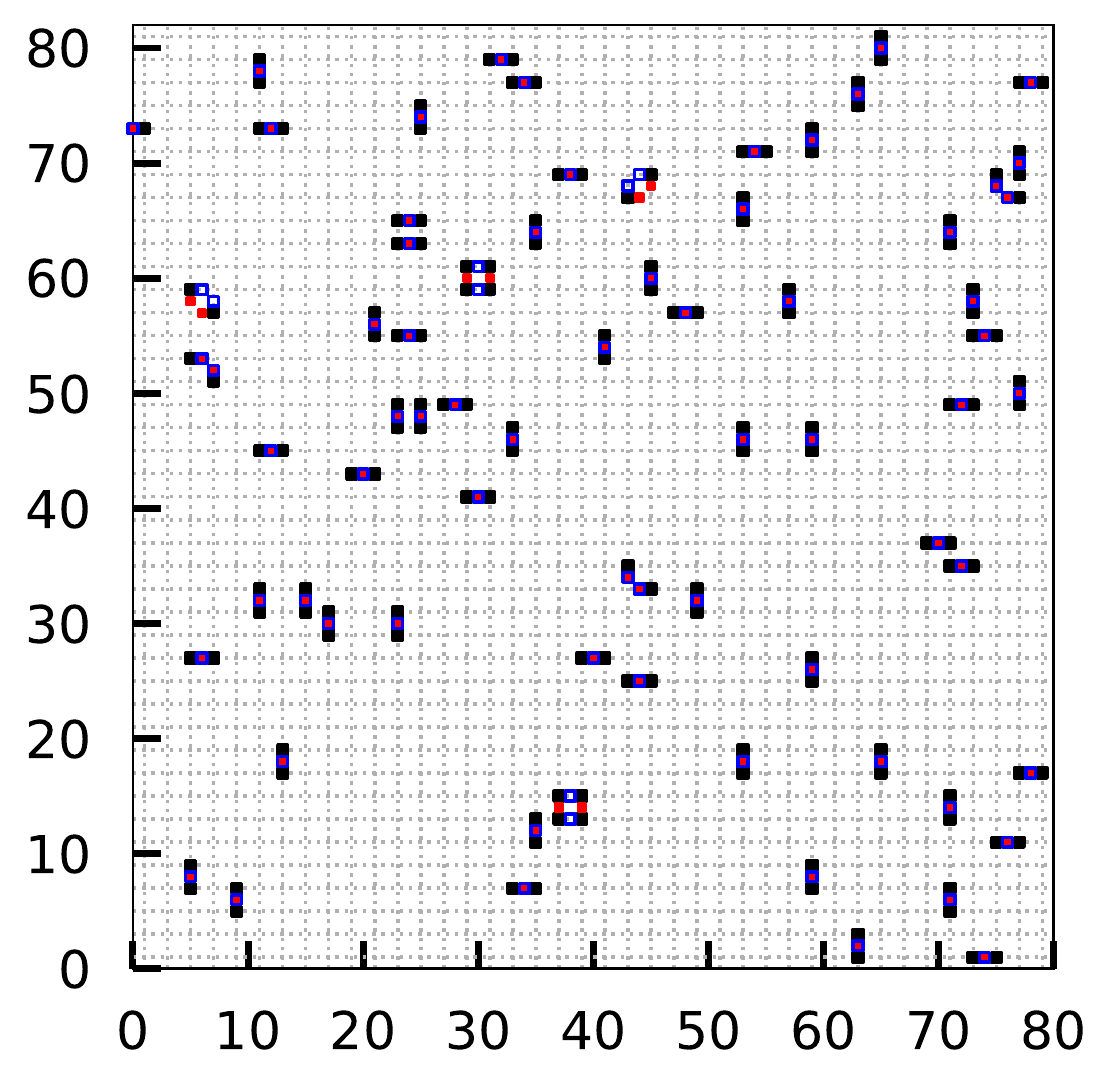}
\caption{\label{fig:sample2}Error correction result in the DA decoder in sample 2.
The filled small squares (red) indicate the actual Z errors, and the open squares (blue) indicate the corrected errors.
Most of the corrected and actual errors coincide, and the rest of them form trivial loops. 
They perfectly reproduce the syndrome.}
\label{fig11}
\end{figure}

The decoding result of the sample 2 for a larger number of qubits with a smaller physical error rate is shown in Fig.~\ref{fig11}.
We can also confirm that in this example 
the syndrome constraint solution is obtained
because the corrected and the actual errors coincide with each other or form trivial loops as in the previous example.
The total energy is the same as the value calculated in the actual errors, and the solution again corresponds to one of the most likely errors.

\section{Characteristics of the DA decoder}
\label{sec:characteristics}

The differences in behavior between the MWPM and DA decoders are discussed here.
In principle, the MWPM decoder always provides a solution with the minimum-distance errors.
In terms of Eq.~(\ref{eq1}), the converged state can be regarded as the global energy minimum (ground state) of the system.
On the other hand, our DA analyses reveals some differences from the MWPM decoder as will be shown below. 

First, we show in Fig.~\ref{fig7} the probability that the solutions obtained with the MWPM and DA decoders are the ground states of the systems. 
\begin{figure}[h]
\includegraphics[width=8.6cm, keepaspectratio]{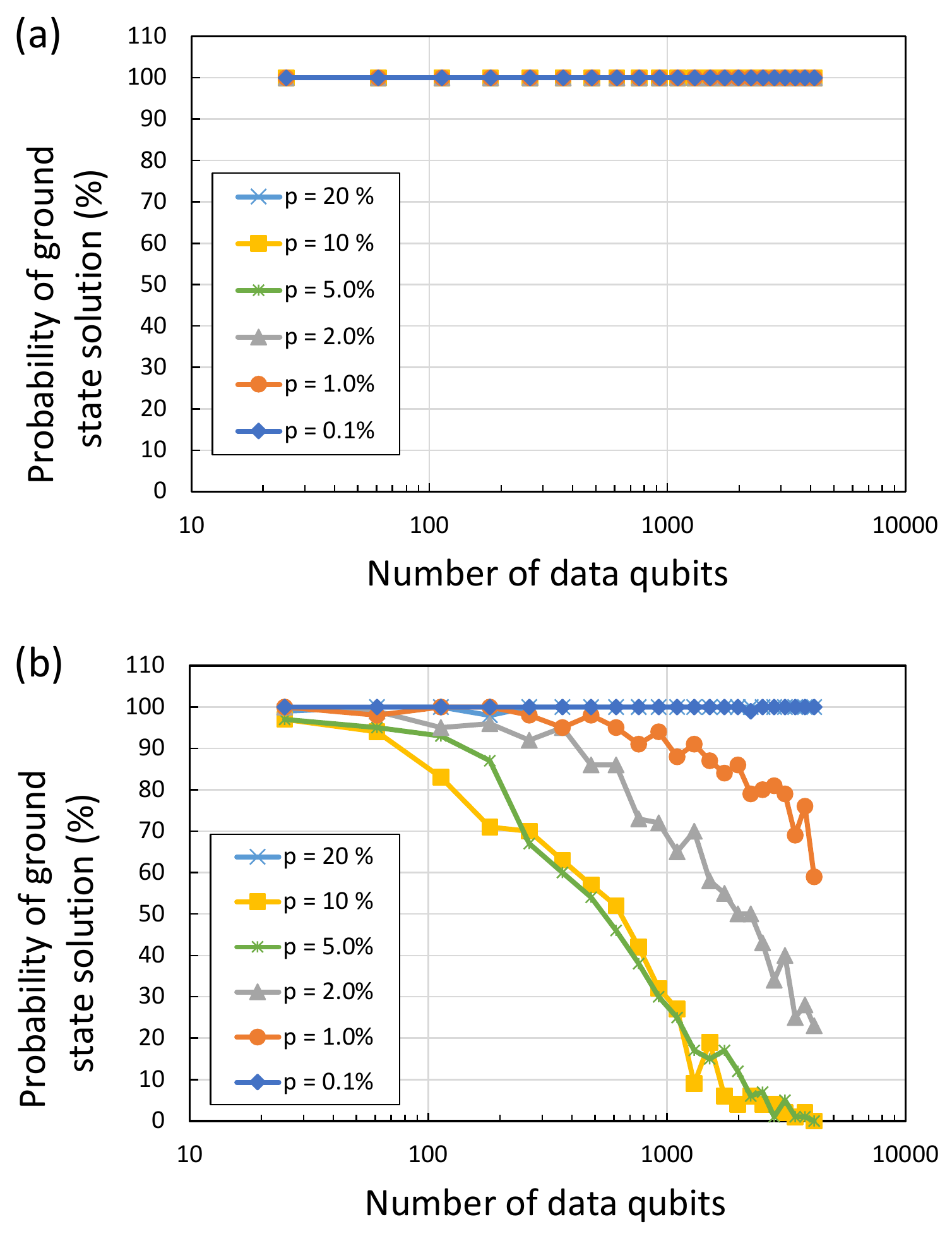}
\caption{\label{fig:ground}Probability of the occurrence of ground states with (a) the MWPM decoder and (b) the DA decoder.
The solutions are always in the ground states for MWPM, whereas the probability of getting ground states for DA decreases with increasing number of qubits.}
\label{fig7}
\end{figure}
For the MWPM decoder, it is obvious that all the solutions are in the ground state, as shown in Fig.~\ref{fig7}(a).
However, for the DA decoder, some of the solutions are in local energy minima (excited states), as shown in Fig.~\ref{fig7}(b).
The probability depends on the number of qubits and the error rate. In particular, when $p$ is 5\% or 10\%, almost all the solutions of the DA decoder are in the excited states for more than 4000 qubits.

\begin{figure}[h]
\includegraphics[width=8.8cm, keepaspectratio]{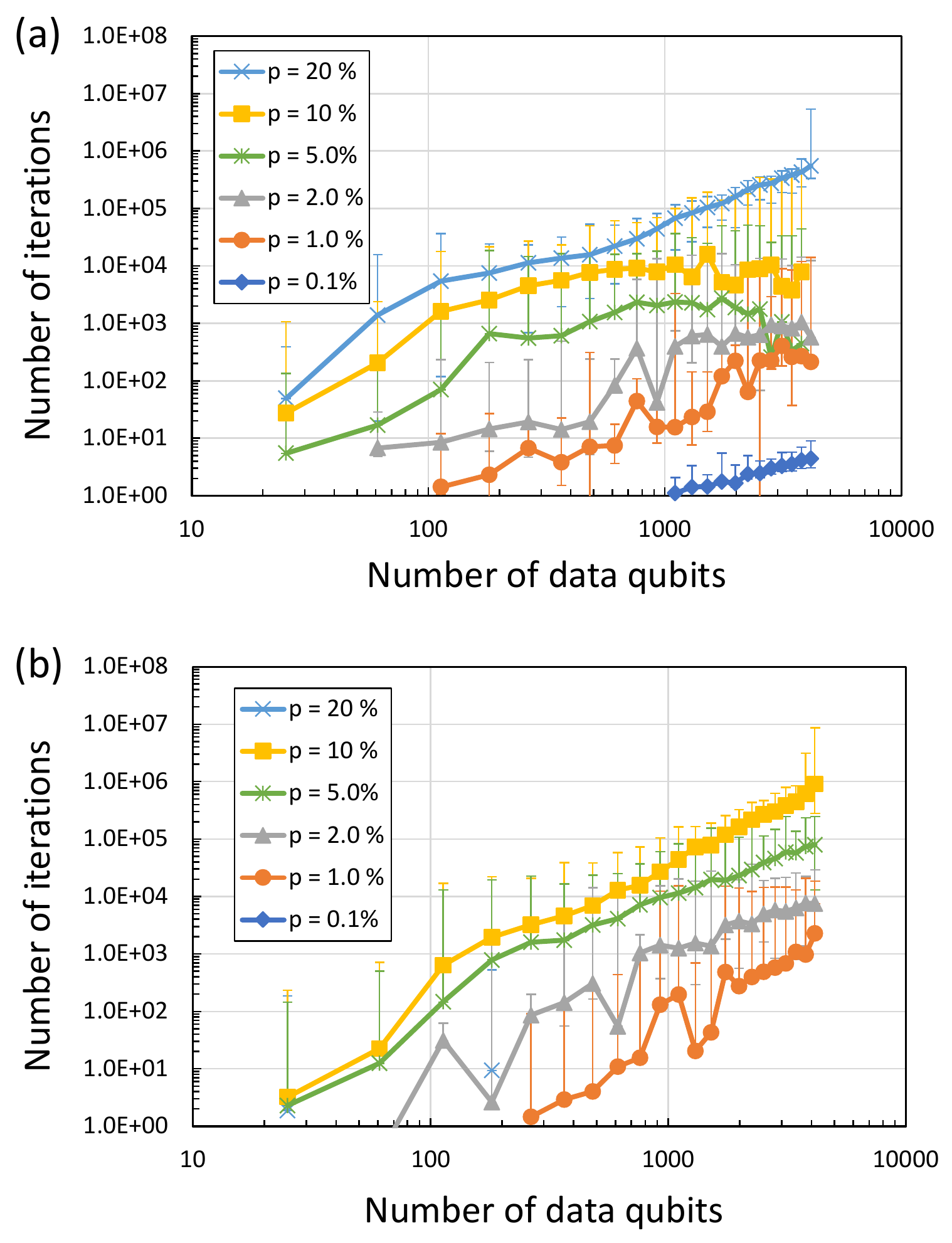}
\caption{\label{fig:geiter}Average numbers of iterations of the DA decoder for each type of convergence.
The results in Fig.~\ref{fig4}(b) are divided into two types. (a) Convergence to the ground states. (b) Convergence into the excited states.}
\label{fig8}
\end{figure}

Since the converged states include both the ground and excited states, we then derived the numbers of iterations for each type of convergence separately.
Figure.~\ref{fig8} shows that the numbers of iterations for the excited states are greater than those for the ground states.
This means that depending on the error pattern, we may obtain a ground state immediately, or we obtain only an excited state after long search.

\begin{figure}[h]
\includegraphics[scale=0.7]{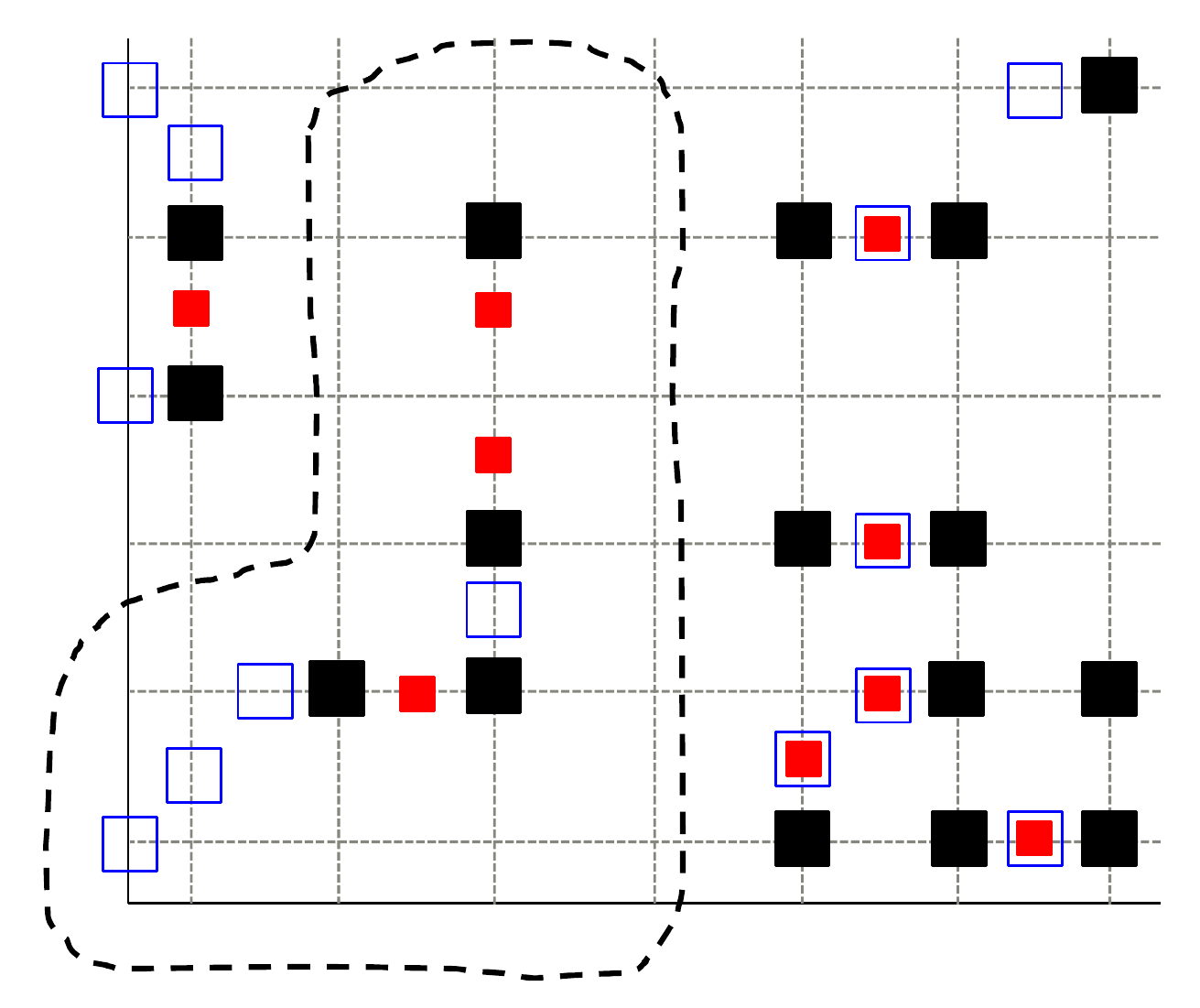}
\caption{\label{fig:chain}Example of a solution breaking the syndrome constraint for $d$ = 32.
A long broken chain made of actual (red filled squares) and corrected (blue open squares) errors is enclosed by a dashed line.}
\label{fig9}
\end{figure}

In order to further investigate the error pattern which causes such long search,
we performed decoding where the upper limit of the number of iterations is set to a relatively low value of 100,000.
All the patterns with which syndrome constraint solutions are not obtained are analyzed for $d$ = 32 and $p$ = 10\%.
We found that the cause of the constraint break was the percolation of the actual and corrected errors forming a long open chain, where the syndrome conditions are not satisfied at its boundary.
An example of the syndrome constraint break is shown in Fig.~\ref{fig9}.
The chain consisting of the actual and corrected errors can be found in the area enclosed by the dashed line, and
this chain causes an syndrome constraint break.
It was also found that, even in the presence of such a chain, an excited state can be obtained finally by further increasing the number of iterations.

Based on the above analyses, the characteristics of the DA decoder are described as follows. When $p$ is sufficiently small, only short chains appear.
In this case, the states rapidly converge to ground states with a small number of iterations.
When $p$ becomes larger and the number of qubits increases, however, the states tend to converge to excited states gradually with some more iterations.
The stochastic approach of DA allows to converge to the excited states in relatively short calculation times and
always keep the syndrome constraint,
even when it is difficult to reach the ground states. 

We should note here that, actual quantum computers are required to operate at a physical error rate sufficiently small, say 1/10 of the error threshold
, in order to reduce the logical error rate sufficiently.
Under such a small physical error rate, though we cannot ignore rare cases like the error patterns described above, the DA decoder can be expected to rapidly give the ground states in practice.
In fact, as shown in Sec. \ref{sec:logical}, the logical error probability is suppressed appropriately, and hence such rare error patterns do not cause false results.
 
\bibliographystyle{apsrev4-1}
\bibliography{bib.bib}

\begin{thebibliography}{38}%
\makeatletter
\providecommand \@ifxundefined [1]{%
 \@ifx{#1\undefined}
}%
\providecommand \@ifnum [1]{%
 \ifnum #1\expandafter \@firstoftwo
 \else \expandafter \@secondoftwo
 \fi
}%
\providecommand \@ifx [1]{%
 \ifx #1\expandafter \@firstoftwo
 \else \expandafter \@secondoftwo
 \fi
}%
\providecommand \natexlab [1]{#1}%
\providecommand \enquote  [1]{``#1''}%
\providecommand \bibnamefont  [1]{#1}%
\providecommand \bibfnamefont [1]{#1}%
\providecommand \citenamefont [1]{#1}%
\providecommand \href@noop [0]{\@secondoftwo}%
\providecommand \href [0]{\begingroup \@sanitize@url \@href}%
\providecommand \@href[1]{\@@startlink{#1}\@@href}%
\providecommand \@@href[1]{\endgroup#1\@@endlink}%
\providecommand \@sanitize@url [0]{\catcode `\\12\catcode `\$12\catcode
  `\&12\catcode `\#12\catcode `\^12\catcode `\_12\catcode `\%12\relax}%
\providecommand \@@startlink[1]{}%
\providecommand \@@endlink[0]{}%
\providecommand \url  [0]{\begingroup\@sanitize@url \@url }%
\providecommand \@url [1]{\endgroup\@href {#1}{\urlprefix }}%
\providecommand \urlprefix  [0]{URL }%
\providecommand \Eprint [0]{\href }%
\providecommand \doibase [0]{http://dx.doi.org/}%
\providecommand \selectlanguage [0]{\@gobble}%
\providecommand \bibinfo  [0]{\@secondoftwo}%
\providecommand \bibfield  [0]{\@secondoftwo}%
\providecommand \translation [1]{[#1]}%
\providecommand \BibitemOpen [0]{}%
\providecommand \bibitemStop [0]{}%
\providecommand \bibitemNoStop [0]{.\EOS\space}%
\providecommand \EOS [0]{\spacefactor3000\relax}%
\providecommand \BibitemShut  [1]{\csname bibitem#1\endcsname}%
\let\auto@bib@innerbib\@empty
\bibitem [{\citenamefont {Shor}(1994)}]{shor1994}%
  \BibitemOpen
  \bibfield  {author} {\bibinfo {author} {\bibfnamefont {P.~W.}\ \bibnamefont
  {Shor}},\ }in\ \href@noop {} {\emph {\bibinfo {booktitle} {Proceedings 35th
  annual symposium on foundations of computer science}}}\ (\bibinfo
  {organization} {Ieee},\ \bibinfo {year} {1994})\ pp.\ \bibinfo {pages}
  {124--134}\BibitemShut {NoStop}%
\bibitem [{\citenamefont {Grover}(1996)}]{grover1996}%
  \BibitemOpen
  \bibfield  {author} {\bibinfo {author} {\bibfnamefont {L.~K.}\ \bibnamefont
  {Grover}},\ }in\ \href@noop {} {\emph {\bibinfo {booktitle} {Proceedings of
  the twenty-eighth annual ACM symposium on Theory of computing}}}\ (\bibinfo
  {year} {1996})\ pp.\ \bibinfo {pages} {212--219}\BibitemShut {NoStop}%
\bibitem [{\citenamefont {Harrow}\ \emph {et~al.}(2009)\citenamefont {Harrow},
  \citenamefont {Hassidim},\ and\ \citenamefont {Lloyd}}]{harrow2009}%
  \BibitemOpen
  \bibfield  {author} {\bibinfo {author} {\bibfnamefont {A.~W.}\ \bibnamefont
  {Harrow}}, \bibinfo {author} {\bibfnamefont {A.}~\bibnamefont {Hassidim}}, \
  and\ \bibinfo {author} {\bibfnamefont {S.}~\bibnamefont {Lloyd}},\
  }\href@noop {} {\bibfield  {journal} {\bibinfo  {journal} {Physical review
  letters}\ }\textbf {\bibinfo {volume} {103}},\ \bibinfo {pages} {150502}
  (\bibinfo {year} {2009})}\BibitemShut {NoStop}%
\bibitem [{\citenamefont {Aspuru-Guzik}\ \emph {et~al.}(2005)\citenamefont
  {Aspuru-Guzik}, \citenamefont {Dutoi}, \citenamefont {Love},\ and\
  \citenamefont {Head-Gordon}}]{aspuru2005}%
  \BibitemOpen
  \bibfield  {author} {\bibinfo {author} {\bibfnamefont {A.}~\bibnamefont
  {Aspuru-Guzik}}, \bibinfo {author} {\bibfnamefont {A.~D.}\ \bibnamefont
  {Dutoi}}, \bibinfo {author} {\bibfnamefont {P.~J.}\ \bibnamefont {Love}}, \
  and\ \bibinfo {author} {\bibfnamefont {M.}~\bibnamefont {Head-Gordon}},\
  }\href@noop {} {\bibfield  {journal} {\bibinfo  {journal} {Science}\ }\textbf
  {\bibinfo {volume} {309}},\ \bibinfo {pages} {1704} (\bibinfo {year}
  {2005})}\BibitemShut {NoStop}%
\bibitem [{\citenamefont {Nielsen}\ and\ \citenamefont
  {Chuang}(2002)}]{nielsen2002}%
  \BibitemOpen
  \bibfield  {author} {\bibinfo {author} {\bibfnamefont {M.~A.}\ \bibnamefont
  {Nielsen}}\ and\ \bibinfo {author} {\bibfnamefont {I.}~\bibnamefont
  {Chuang}},\ }\href@noop {} {\enquote {\bibinfo {title} {Quantum computation
  and quantum information},}\ } (\bibinfo {year} {2002})\BibitemShut {NoStop}%
\bibitem [{\citenamefont {Reiher}\ \emph {et~al.}(2017)\citenamefont {Reiher},
  \citenamefont {Wiebe}, \citenamefont {Svore}, \citenamefont {Wecker},\ and\
  \citenamefont {Troyer}}]{reiher2017}%
  \BibitemOpen
  \bibfield  {author} {\bibinfo {author} {\bibfnamefont {M.}~\bibnamefont
  {Reiher}}, \bibinfo {author} {\bibfnamefont {N.}~\bibnamefont {Wiebe}},
  \bibinfo {author} {\bibfnamefont {K.~M.}\ \bibnamefont {Svore}}, \bibinfo
  {author} {\bibfnamefont {D.}~\bibnamefont {Wecker}}, \ and\ \bibinfo {author}
  {\bibfnamefont {M.}~\bibnamefont {Troyer}},\ }\href@noop {} {\bibfield
  {journal} {\bibinfo  {journal} {Proceedings of the National Academy of
  Sciences}\ }\textbf {\bibinfo {volume} {114}},\ \bibinfo {pages} {7555}
  (\bibinfo {year} {2017})}\BibitemShut {NoStop}%
\bibitem [{\citenamefont {Gidney}\ and\ \citenamefont
  {Eker{\aa}}(2021)}]{gidney2021}%
  \BibitemOpen
  \bibfield  {author} {\bibinfo {author} {\bibfnamefont {C.}~\bibnamefont
  {Gidney}}\ and\ \bibinfo {author} {\bibfnamefont {M.}~\bibnamefont
  {Eker{\aa}}},\ }\href@noop {} {\bibfield  {journal} {\bibinfo  {journal}
  {Quantum}\ }\textbf {\bibinfo {volume} {5}},\ \bibinfo {pages} {433}
  (\bibinfo {year} {2021})}\BibitemShut {NoStop}%
\bibitem [{\citenamefont {Kelly}\ \emph {et~al.}(2015)\citenamefont {Kelly},
  \citenamefont {Barends}, \citenamefont {Fowler}, \citenamefont {Megrant},
  \citenamefont {Jeffrey}, \citenamefont {White}, \citenamefont {Sank},
  \citenamefont {Mutus}, \citenamefont {Campbell}, \citenamefont {Chen} \emph
  {et~al.}}]{kelly2015state}%
  \BibitemOpen
  \bibfield  {author} {\bibinfo {author} {\bibfnamefont {J.}~\bibnamefont
  {Kelly}}, \bibinfo {author} {\bibfnamefont {R.}~\bibnamefont {Barends}},
  \bibinfo {author} {\bibfnamefont {A.~G.}\ \bibnamefont {Fowler}}, \bibinfo
  {author} {\bibfnamefont {A.}~\bibnamefont {Megrant}}, \bibinfo {author}
  {\bibfnamefont {E.}~\bibnamefont {Jeffrey}}, \bibinfo {author} {\bibfnamefont
  {T.~C.}\ \bibnamefont {White}}, \bibinfo {author} {\bibfnamefont
  {D.}~\bibnamefont {Sank}}, \bibinfo {author} {\bibfnamefont {J.~Y.}\
  \bibnamefont {Mutus}}, \bibinfo {author} {\bibfnamefont {B.}~\bibnamefont
  {Campbell}}, \bibinfo {author} {\bibfnamefont {Y.}~\bibnamefont {Chen}},
  \emph {et~al.},\ }\href@noop {} {\bibfield  {journal} {\bibinfo  {journal}
  {Nature}\ }\textbf {\bibinfo {volume} {519}},\ \bibinfo {pages} {66}
  (\bibinfo {year} {2015})}\BibitemShut {NoStop}%
\bibitem [{\citenamefont {Andersen}\ \emph {et~al.}(2020)\citenamefont
  {Andersen}, \citenamefont {Remm}, \citenamefont {Lazar}, \citenamefont
  {Krinner}, \citenamefont {Lacroix}, \citenamefont {Norris}, \citenamefont
  {Gabureac}, \citenamefont {Eichler},\ and\ \citenamefont
  {Wallraff}}]{andersen2020repeated}%
  \BibitemOpen
  \bibfield  {author} {\bibinfo {author} {\bibfnamefont {C.~K.}\ \bibnamefont
  {Andersen}}, \bibinfo {author} {\bibfnamefont {A.}~\bibnamefont {Remm}},
  \bibinfo {author} {\bibfnamefont {S.}~\bibnamefont {Lazar}}, \bibinfo
  {author} {\bibfnamefont {S.}~\bibnamefont {Krinner}}, \bibinfo {author}
  {\bibfnamefont {N.}~\bibnamefont {Lacroix}}, \bibinfo {author} {\bibfnamefont
  {G.~J.}\ \bibnamefont {Norris}}, \bibinfo {author} {\bibfnamefont
  {M.}~\bibnamefont {Gabureac}}, \bibinfo {author} {\bibfnamefont
  {C.}~\bibnamefont {Eichler}}, \ and\ \bibinfo {author} {\bibfnamefont
  {A.}~\bibnamefont {Wallraff}},\ }\href@noop {} {\bibfield  {journal}
  {\bibinfo  {journal} {Nature Physics}\ }\textbf {\bibinfo {volume} {16}},\
  \bibinfo {pages} {875} (\bibinfo {year} {2020})}\BibitemShut {NoStop}%
\bibitem [{\citenamefont {Egan}\ \emph {et~al.}(2021)\citenamefont {Egan},
  \citenamefont {Debroy}, \citenamefont {Noel}, \citenamefont {Risinger},
  \citenamefont {Zhu}, \citenamefont {Biswas}, \citenamefont {Newman},
  \citenamefont {Li}, \citenamefont {Brown}, \citenamefont {Cetina} \emph
  {et~al.}}]{egan2021fault}%
  \BibitemOpen
  \bibfield  {author} {\bibinfo {author} {\bibfnamefont {L.}~\bibnamefont
  {Egan}}, \bibinfo {author} {\bibfnamefont {D.~M.}\ \bibnamefont {Debroy}},
  \bibinfo {author} {\bibfnamefont {C.}~\bibnamefont {Noel}}, \bibinfo {author}
  {\bibfnamefont {A.}~\bibnamefont {Risinger}}, \bibinfo {author}
  {\bibfnamefont {D.}~\bibnamefont {Zhu}}, \bibinfo {author} {\bibfnamefont
  {D.}~\bibnamefont {Biswas}}, \bibinfo {author} {\bibfnamefont
  {M.}~\bibnamefont {Newman}}, \bibinfo {author} {\bibfnamefont
  {M.}~\bibnamefont {Li}}, \bibinfo {author} {\bibfnamefont {K.~R.}\
  \bibnamefont {Brown}}, \bibinfo {author} {\bibfnamefont {M.}~\bibnamefont
  {Cetina}},  \emph {et~al.},\ }\href@noop {} {\bibfield  {journal} {\bibinfo
  {journal} {Nature}\ }\textbf {\bibinfo {volume} {598}},\ \bibinfo {pages}
  {281} (\bibinfo {year} {2021})}\BibitemShut {NoStop}%
\bibitem [{\citenamefont {AI}(2021)}]{ai2021exponential}%
  \BibitemOpen
  \bibfield  {author} {\bibinfo {author} {\bibfnamefont {G.~Q.}\ \bibnamefont
  {AI}},\ }\href@noop {} {\bibfield  {journal} {\bibinfo  {journal} {Nature}\
  }\textbf {\bibinfo {volume} {595}},\ \bibinfo {pages} {383} (\bibinfo {year}
  {2021})}\BibitemShut {NoStop}%
\bibitem [{\citenamefont {Krinner}\ \emph {et~al.}(2021)\citenamefont
  {Krinner}, \citenamefont {Lacroix}, \citenamefont {Remm}, \citenamefont
  {Paolo}, \citenamefont {Genois}, \citenamefont {Leroux}, \citenamefont
  {Hellings}, \citenamefont {Lazar}, \citenamefont {Swiadek}, \citenamefont
  {Herrmann}, \citenamefont {Norris}, \citenamefont {Andersen}, \citenamefont
  {Müller}, \citenamefont {Blais}, \citenamefont {Eichler},\ and\
  \citenamefont {Wallraff}}]{krinner2021realizing}%
  \BibitemOpen
  \bibfield  {author} {\bibinfo {author} {\bibfnamefont {S.}~\bibnamefont
  {Krinner}}, \bibinfo {author} {\bibfnamefont {N.}~\bibnamefont {Lacroix}},
  \bibinfo {author} {\bibfnamefont {A.}~\bibnamefont {Remm}}, \bibinfo {author}
  {\bibfnamefont {A.~D.}\ \bibnamefont {Paolo}}, \bibinfo {author}
  {\bibfnamefont {E.}~\bibnamefont {Genois}}, \bibinfo {author} {\bibfnamefont
  {C.}~\bibnamefont {Leroux}}, \bibinfo {author} {\bibfnamefont
  {C.}~\bibnamefont {Hellings}}, \bibinfo {author} {\bibfnamefont
  {S.}~\bibnamefont {Lazar}}, \bibinfo {author} {\bibfnamefont
  {F.}~\bibnamefont {Swiadek}}, \bibinfo {author} {\bibfnamefont
  {J.}~\bibnamefont {Herrmann}}, \bibinfo {author} {\bibfnamefont {G.~J.}\
  \bibnamefont {Norris}}, \bibinfo {author} {\bibfnamefont {C.~K.}\
  \bibnamefont {Andersen}}, \bibinfo {author} {\bibfnamefont {M.}~\bibnamefont
  {Müller}}, \bibinfo {author} {\bibfnamefont {A.}~\bibnamefont {Blais}},
  \bibinfo {author} {\bibfnamefont {C.}~\bibnamefont {Eichler}}, \ and\
  \bibinfo {author} {\bibfnamefont {A.}~\bibnamefont {Wallraff}},\ }\href@noop
  {} {\enquote {\bibinfo {title} {Realizing repeated quantum error correction
  in a distance-three surface code},}\ } (\bibinfo {year} {2021}),\ \Eprint
  {http://arxiv.org/abs/2112.03708} {arXiv:2112.03708 [quant-ph]} \BibitemShut
  {NoStop}%
\bibitem [{\citenamefont {Das}\ \emph {et~al.}(2020)\citenamefont {Das},
  \citenamefont {Pattison}, \citenamefont {Manne}, \citenamefont {Carmean},
  \citenamefont {Svore}, \citenamefont {Qureshi},\ and\ \citenamefont
  {Delfosse}}]{das2020scalable}%
  \BibitemOpen
  \bibfield  {author} {\bibinfo {author} {\bibfnamefont {P.}~\bibnamefont
  {Das}}, \bibinfo {author} {\bibfnamefont {C.~A.}\ \bibnamefont {Pattison}},
  \bibinfo {author} {\bibfnamefont {S.}~\bibnamefont {Manne}}, \bibinfo
  {author} {\bibfnamefont {D.}~\bibnamefont {Carmean}}, \bibinfo {author}
  {\bibfnamefont {K.}~\bibnamefont {Svore}}, \bibinfo {author} {\bibfnamefont
  {M.}~\bibnamefont {Qureshi}}, \ and\ \bibinfo {author} {\bibfnamefont
  {N.}~\bibnamefont {Delfosse}},\ }\href@noop {} {\bibfield  {journal}
  {\bibinfo  {journal} {arXiv preprint arXiv:2001.06598}\ } (\bibinfo {year}
  {2020})}\BibitemShut {NoStop}%
\bibitem [{\citenamefont {Holmes}\ \emph {et~al.}(2020)\citenamefont {Holmes},
  \citenamefont {Jokar}, \citenamefont {Pasandi}, \citenamefont {Ding},
  \citenamefont {Pedram},\ and\ \citenamefont {Chong}}]{Holmes20}%
  \BibitemOpen
  \bibfield  {author} {\bibinfo {author} {\bibfnamefont {A.}~\bibnamefont
  {Holmes}}, \bibinfo {author} {\bibfnamefont {M.~R.}\ \bibnamefont {Jokar}},
  \bibinfo {author} {\bibfnamefont {G.}~\bibnamefont {Pasandi}}, \bibinfo
  {author} {\bibfnamefont {Y.}~\bibnamefont {Ding}}, \bibinfo {author}
  {\bibfnamefont {M.}~\bibnamefont {Pedram}}, \ and\ \bibinfo {author}
  {\bibfnamefont {F.~T.}\ \bibnamefont {Chong}},\ }\href@noop {} {\bibfield
  {journal} {\bibinfo  {journal} {2020 ACM/IEEE 47th Annual International
  Symposium on Computer Architecture (ISCA)}\ ,\ \bibinfo {pages} {556}}
  (\bibinfo {year} {2020})}\BibitemShut {NoStop}%
\bibitem [{\citenamefont {Ueno}\ \emph {et~al.}(2021)\citenamefont {Ueno},
  \citenamefont {Kondo}, \citenamefont {Tanaka}, \citenamefont {Suzuki},\ and\
  \citenamefont {Tabuchi}}]{ueno2021qecool}%
  \BibitemOpen
  \bibfield  {author} {\bibinfo {author} {\bibfnamefont {Y.}~\bibnamefont
  {Ueno}}, \bibinfo {author} {\bibfnamefont {M.}~\bibnamefont {Kondo}},
  \bibinfo {author} {\bibfnamefont {M.}~\bibnamefont {Tanaka}}, \bibinfo
  {author} {\bibfnamefont {Y.}~\bibnamefont {Suzuki}}, \ and\ \bibinfo {author}
  {\bibfnamefont {Y.}~\bibnamefont {Tabuchi}},\ }\href@noop {} {\bibfield
  {journal} {\bibinfo  {journal} {arXiv preprint arXiv:2103.14209}\ } (\bibinfo
  {year} {2021})}\BibitemShut {NoStop}%
\bibitem [{\citenamefont {Fujii}\ \emph {et~al.}(2014)\citenamefont {Fujii},
  \citenamefont {Negoro}, \citenamefont {Imoto},\ and\ \citenamefont
  {Kitagawa}}]{fujii2014}%
  \BibitemOpen
  \bibfield  {author} {\bibinfo {author} {\bibfnamefont {K.}~\bibnamefont
  {Fujii}}, \bibinfo {author} {\bibfnamefont {M.}~\bibnamefont {Negoro}},
  \bibinfo {author} {\bibfnamefont {N.}~\bibnamefont {Imoto}}, \ and\ \bibinfo
  {author} {\bibfnamefont {M.}~\bibnamefont {Kitagawa}},\ }\href@noop {}
  {\bibfield  {journal} {\bibinfo  {journal} {Physical Review X}\ }\textbf
  {\bibinfo {volume} {4}},\ \bibinfo {pages} {041039} (\bibinfo {year}
  {2014})}\BibitemShut {NoStop}%
\bibitem [{\citenamefont {M.~Herold}\ and\ \citenamefont
  {Kastoryano}(2015)}]{Herold2015}%
  \BibitemOpen
  \bibfield  {author} {\bibinfo {author} {\bibfnamefont {J.~E.}\ \bibnamefont
  {M.~Herold}, \bibfnamefont {E.~T.~Campbell}}\ and\ \bibinfo {author}
  {\bibfnamefont {M.~J.}\ \bibnamefont {Kastoryano}},\ }\href@noop {}
  {\bibfield  {journal} {\bibinfo  {journal} {npj Quantum Inf}\ }\textbf
  {\bibinfo {volume} {1}},\ \bibinfo {pages} {15010} (\bibinfo {year}
  {2015})}\BibitemShut {NoStop}%
\bibitem [{\citenamefont {M.~Herold}\ and\ \citenamefont
  {Eisert}(2017)}]{Herold2017}%
  \BibitemOpen
  \bibfield  {author} {\bibinfo {author} {\bibfnamefont {E.~T.~C.}\
  \bibnamefont {M.~Herold}, \bibfnamefont {M.~J.~Kastoryano}}\ and\ \bibinfo
  {author} {\bibfnamefont {J.}~\bibnamefont {Eisert}},\ }\href@noop {}
  {\bibfield  {journal} {\bibinfo  {journal} {New J. Phys.}\ }\textbf {\bibinfo
  {volume} {19}},\ \bibinfo {pages} {063012} (\bibinfo {year}
  {2017})}\BibitemShut {NoStop}%
\bibitem [{\citenamefont {Kubica}\ and\ \citenamefont
  {Preskill}(2018)}]{Kubica2018}%
  \BibitemOpen
  \bibfield  {author} {\bibinfo {author} {\bibfnamefont {A.}~\bibnamefont
  {Kubica}}\ and\ \bibinfo {author} {\bibfnamefont {J.}~\bibnamefont
  {Preskill}},\ }\href@noop {} {\bibfield  {journal} {\bibinfo  {journal}
  {arXiv:1809.10145}\ } (\bibinfo {year} {2018})}\BibitemShut {NoStop}%
\bibitem [{\citenamefont {Sao}\ \emph {et~al.}(2019)\citenamefont {Sao},
  \citenamefont {Watanabe}, \citenamefont {Musha},\ and\ \citenamefont
  {Utsunomiya}}]{Sao19}%
  \BibitemOpen
  \bibfield  {author} {\bibinfo {author} {\bibfnamefont {M.}~\bibnamefont
  {Sao}}, \bibinfo {author} {\bibfnamefont {H.}~\bibnamefont {Watanabe}},
  \bibinfo {author} {\bibfnamefont {Y.}~\bibnamefont {Musha}}, \ and\ \bibinfo
  {author} {\bibfnamefont {A.}~\bibnamefont {Utsunomiya}},\ }\href@noop {}
  {\bibfield  {journal} {\bibinfo  {journal} {FUJITSU SCIENTIFIC \& TECHNICAL
  JOURNAL}\ }\textbf {\bibinfo {volume} {55}},\ \bibinfo {pages} {45} (\bibinfo
  {year} {2019})}\BibitemShut {NoStop}%
\bibitem [{\citenamefont {Matsubara}\ \emph {et~al.}(2020)\citenamefont
  {Matsubara}, \citenamefont {Takatsu}, \citenamefont {Miyazawa}, \citenamefont
  {Shibasaki}, \citenamefont {Watanabe}, \citenamefont {Takemoto},\ and\
  \citenamefont {Tamura}}]{Matsubara20}%
  \BibitemOpen
  \bibfield  {author} {\bibinfo {author} {\bibfnamefont {S.}~\bibnamefont
  {Matsubara}}, \bibinfo {author} {\bibfnamefont {M.}~\bibnamefont {Takatsu}},
  \bibinfo {author} {\bibfnamefont {T.}~\bibnamefont {Miyazawa}}, \bibinfo
  {author} {\bibfnamefont {T.}~\bibnamefont {Shibasaki}}, \bibinfo {author}
  {\bibfnamefont {Y.}~\bibnamefont {Watanabe}}, \bibinfo {author}
  {\bibfnamefont {K.}~\bibnamefont {Takemoto}}, \ and\ \bibinfo {author}
  {\bibfnamefont {H.}~\bibnamefont {Tamura}},\ }\href@noop {} {\bibfield
  {journal} {\bibinfo  {journal} {25th Asia and South Pacific Design Automation
  Conference (ASP-DAC)}\ } (\bibinfo {year} {2020})}\BibitemShut {NoStop}%
\bibitem [{\citenamefont {Aramon}\ \emph {et~al.}(2019)\citenamefont {Aramon},
  \citenamefont {Rosenberg}, \citenamefont {Valiante}, \citenamefont
  {Miyazawa}, \citenamefont {Tamura}, ,\ and\ \citenamefont
  {Katzgraber}}]{Aramon19}%
  \BibitemOpen
  \bibfield  {author} {\bibinfo {author} {\bibfnamefont {M.}~\bibnamefont
  {Aramon}}, \bibinfo {author} {\bibfnamefont {G.}~\bibnamefont {Rosenberg}},
  \bibinfo {author} {\bibfnamefont {E.}~\bibnamefont {Valiante}}, \bibinfo
  {author} {\bibfnamefont {T.}~\bibnamefont {Miyazawa}}, \bibinfo {author}
  {\bibfnamefont {H.}~\bibnamefont {Tamura}}, , \ and\ \bibinfo {author}
  {\bibfnamefont {H.~G.}\ \bibnamefont {Katzgraber}},\ }\href@noop {}
  {\bibfield  {journal} {\bibinfo  {journal} {Frontiers in Physics}\ }\textbf
  {\bibinfo {volume} {7}} (\bibinfo {year} {2019})}\BibitemShut {NoStop}%
\bibitem [{DA()}]{DA}%
  \BibitemOpen
  \href@noop {} {\enquote {\bibinfo {title} {{Official website of Fujitsu's
  Digital Annealer}},}\ }\bibinfo {howpublished}
  {\url{https://www.fujitsu.com/global/services/business-services/digital-annealer/}}\BibitemShut
  {NoStop}%
\bibitem [{\citenamefont {Kirkpatrick}(1983)}]{Kirkpatrick83}%
  \BibitemOpen
  \bibfield  {author} {\bibinfo {author} {\bibfnamefont {S.}~\bibnamefont
  {Kirkpatrick}},\ }\href@noop {} {\bibfield  {journal} {\bibinfo  {journal}
  {Science}\ }\textbf {\bibinfo {volume} {220}},\ \bibinfo {pages} {671}
  (\bibinfo {year} {1983})}\BibitemShut {NoStop}%
\bibitem [{\citenamefont {Kadowaki}\ and\ \citenamefont
  {Nishimori}(1998)}]{Kadowaki98}%
  \BibitemOpen
  \bibfield  {author} {\bibinfo {author} {\bibfnamefont {T.}~\bibnamefont
  {Kadowaki}}\ and\ \bibinfo {author} {\bibfnamefont {H.}~\bibnamefont
  {Nishimori}},\ }\href@noop {} {\bibfield  {journal} {\bibinfo  {journal}
  {Phys. Rev. E}\ }\textbf {\bibinfo {volume} {58}},\ \bibinfo {pages} {5355}
  (\bibinfo {year} {1998})}\BibitemShut {NoStop}%
\bibitem [{\citenamefont {Hukushima}\ and\ \citenamefont
  {Nemoto}(1996)}]{Hukushima96}%
  \BibitemOpen
  \bibfield  {author} {\bibinfo {author} {\bibfnamefont {K.}~\bibnamefont
  {Hukushima}}\ and\ \bibinfo {author} {\bibfnamefont {K.}~\bibnamefont
  {Nemoto}},\ }\href@noop {} {\bibfield  {journal} {\bibinfo  {journal} {J.
  Phys. Soc. Jpn.}\ }\textbf {\bibinfo {volume} {65}},\ \bibinfo {pages} {1604}
  (\bibinfo {year} {1996})}\BibitemShut {NoStop}%
\bibitem [{\citenamefont {Edmonds}(1965)}]{Edmonds65}%
  \BibitemOpen
  \bibfield  {author} {\bibinfo {author} {\bibfnamefont {J.}~\bibnamefont
  {Edmonds}},\ }\href@noop {} {\bibfield  {journal} {\bibinfo  {journal}
  {Canadian Journal of mathematics}\ }\textbf {\bibinfo {volume} {17.3}},\
  \bibinfo {pages} {449} (\bibinfo {year} {1965})}\BibitemShut {NoStop}%
\bibitem [{\citenamefont {Galil}(1986)}]{Galil86}%
  \BibitemOpen
  \bibfield  {author} {\bibinfo {author} {\bibfnamefont {Z.}~\bibnamefont
  {Galil}},\ }\href@noop {} {\bibfield  {journal} {\bibinfo  {journal} {ACM
  Computing Surveys}\ }\textbf {\bibinfo {volume} {18}},\ \bibinfo {pages} {23}
  (\bibinfo {year} {1986})}\BibitemShut {NoStop}%
\bibitem [{\citenamefont {Delfosse}\ and\ \citenamefont
  {Nickerson}(2017)}]{Delfosse17}%
  \BibitemOpen
  \bibfield  {author} {\bibinfo {author} {\bibfnamefont {N.}~\bibnamefont
  {Delfosse}}\ and\ \bibinfo {author} {\bibfnamefont {N.}~\bibnamefont
  {Nickerson}},\ }\href@noop {} {\bibfield  {journal} {\bibinfo  {journal}
  {arXiv:1709.06218v1}\ } (\bibinfo {year} {2017})}\BibitemShut {NoStop}%
\bibitem [{\citenamefont {Kitaev}(2003)}]{Kitaev2003}%
  \BibitemOpen
  \bibfield  {author} {\bibinfo {author} {\bibfnamefont {A.~Y.}\ \bibnamefont
  {Kitaev}},\ }\href@noop {} {\bibfield  {journal} {\bibinfo  {journal} {Annals
  of Physics}\ }\textbf {\bibinfo {volume} {303}},\ \bibinfo {pages} {2}
  (\bibinfo {year} {2003})}\BibitemShut {NoStop}%
\bibitem [{\citenamefont {Bravyi}\ and\ \citenamefont
  {Kitaev}(1998)}]{Bravyi1998}%
  \BibitemOpen
  \bibfield  {author} {\bibinfo {author} {\bibfnamefont {S.~B.}\ \bibnamefont
  {Bravyi}}\ and\ \bibinfo {author} {\bibfnamefont {A.~Y.}\ \bibnamefont
  {Kitaev}},\ }\href@noop {} {\bibfield  {journal} {\bibinfo  {journal}
  {arXiv}\ } (\bibinfo {year} {1998})}\BibitemShut {NoStop}%
\bibitem [{\citenamefont {Fowler}\ \emph {et~al.}(2012)\citenamefont {Fowler},
  \citenamefont {Mariantoni}, \citenamefont {Martinis},\ and\ \citenamefont
  {Cleland}}]{Fowler12}%
  \BibitemOpen
  \bibfield  {author} {\bibinfo {author} {\bibfnamefont {A.}~\bibnamefont
  {Fowler}}, \bibinfo {author} {\bibfnamefont {M.}~\bibnamefont {Mariantoni}},
  \bibinfo {author} {\bibfnamefont {J.}~\bibnamefont {Martinis}}, \ and\
  \bibinfo {author} {\bibfnamefont {A.}~\bibnamefont {Cleland}},\ }\href@noop
  {} {\bibfield  {journal} {\bibinfo  {journal} {Phys. Rev. A}\ }\textbf
  {\bibinfo {volume} {86}},\ \bibinfo {pages} {032324} (\bibinfo {year}
  {2012})}\BibitemShut {NoStop}%
\bibitem [{\citenamefont {Ishikawa}(2011)}]{Ishikawa11}%
  \BibitemOpen
  \bibfield  {author} {\bibinfo {author} {\bibfnamefont {H.}~\bibnamefont
  {Ishikawa}},\ }\href@noop {} {\bibfield  {journal} {\bibinfo  {journal} {IEEE
  Transactions on Pattern Analysis and Machine Intelligence}\ }\textbf
  {\bibinfo {volume} {33}},\ \bibinfo {pages} {1234} (\bibinfo {year}
  {2011})}\BibitemShut {NoStop}%
\bibitem [{\citenamefont {Xia}\ \emph {et~al.}(2018)\citenamefont {Xia},
  \citenamefont {Bian},\ and\ \citenamefont {Kais}}]{Xia18}%
  \BibitemOpen
  \bibfield  {author} {\bibinfo {author} {\bibfnamefont {R.}~\bibnamefont
  {Xia}}, \bibinfo {author} {\bibfnamefont {T.}~\bibnamefont {Bian}}, \ and\
  \bibinfo {author} {\bibfnamefont {S.}~\bibnamefont {Kais}},\ }\href@noop {}
  {\bibfield  {journal} {\bibinfo  {journal} {J. Phys. Chem. B}\ }\textbf
  {\bibinfo {volume} {122}},\ \bibinfo {pages} {3384} (\bibinfo {year}
  {2018})}\BibitemShut {NoStop}%
\bibitem [{Net()}]{Networkx}%
  \BibitemOpen
  \href@noop {} {\enquote {\bibinfo {title} {{Official website of Networkx}},}\
  }\bibinfo {howpublished} {\url{https://networkx.org/}}\BibitemShut {NoStop}%
\bibitem [{\citenamefont {Duclos-Cianci}\ and\ \citenamefont
  {Poulin}(2010{\natexlab{a}})}]{Duclos-Cianci2010a}%
  \BibitemOpen
  \bibfield  {author} {\bibinfo {author} {\bibfnamefont {G.}~\bibnamefont
  {Duclos-Cianci}}\ and\ \bibinfo {author} {\bibfnamefont {D.}~\bibnamefont
  {Poulin}},\ }\href {\doibase 10.1109/CIG.2010.5592866} {\bibfield  {journal}
  {\bibinfo  {journal} {IEEE Information Theory Workshop}\ ,\ \bibinfo {pages}
  {1}} (\bibinfo {year} {2010}{\natexlab{a}})}\BibitemShut {NoStop}%
\bibitem [{\citenamefont {Duclos-Cianci}\ and\ \citenamefont
  {Poulin}(2010{\natexlab{b}})}]{Duclos-Cianci2010b}%
  \BibitemOpen
  \bibfield  {author} {\bibinfo {author} {\bibfnamefont {G.}~\bibnamefont
  {Duclos-Cianci}}\ and\ \bibinfo {author} {\bibfnamefont {D.}~\bibnamefont
  {Poulin}},\ }\href@noop {} {\bibfield  {journal} {\bibinfo  {journal} {Phys.
  Rev. Lett.}\ }\textbf {\bibinfo {volume} {104}},\ \bibinfo {pages} {050504}
  (\bibinfo {year} {2010}{\natexlab{b}})}\BibitemShut {NoStop}%
\bibitem [{\citenamefont {Bombin}(2013)}]{Bombin13}%
  \BibitemOpen
  \bibfield  {author} {\bibinfo {author} {\bibfnamefont {H.}~\bibnamefont
  {Bombin}},\ }\href@noop {} {\bibfield  {journal} {\bibinfo  {journal}
  {arXiv:1311.0277v1}\ } (\bibinfo {year} {2013})}\BibitemShut {NoStop}%
\end{thebibliography}%

\end{document}